%% file: ASKAP_SAM_MNRAS.tex
\documentclass[useAMS,usenatbib]{mn2e}

\newcommand{\Msol}{\, \rm M_{\odot}}
\newcommand{\hMsol}{\, h^{-1}{\rm M_{\odot}}}
\newcommand{\hMpc}{\, h^{-1}{\rm Mpc}}
\newcommand{\Mpc}{\, \rm Mpc}

\newcommand{\kms}{\, {\rm km}\, {\rm s}^{-1}}
\newcommand{\aeff}{A_{\rm eff}}
\newcommand{\tsys}{T_{\rm sys}}

\newcommand{\sqd}{\,\rm deg^{2}}

\newcommand{\HI}{{H{\sc i}\,\,}}
\newcommand{\Htwo}{{$\rm H_2\,\,$}}

\usepackage[authoryear]{natbib}
\bibpunct{(}{)}{;}{a}{}{,}
\usepackage{graphicx}
\usepackage{rotating}
\usepackage{epsfig}
\usepackage{color}
\usepackage{float}
\usepackage{fixltx2e}
\usepackage{multirow}
\usepackage{url}

\input{jdefs}

\title[ASKAP surveys]{Predictions for ASKAP Neutral Hydrogen Surveys}
\author[A. R. Duffy et al.]
{Alan R. Duffy$^{1,2}$, Martin J. Meyer$^{1}$, Lister Staveley-Smith$^{1}$, Maksym Bernyk$^{3}$,  \newauthor 
Darren J. Croton$^{3}$, B\"arbel S. Koribalski$^{4}$, Derek Gerstmann$^{1}$, Stefan Westerlund$^{1}$ \\
$^1$International Centre for Radio Astronomy Research, The University of Western Australia, Crawley, WA 6009, Australia \\
$^2$School of Physics, University of Melbourne, Parkville, Victoria 3010, Australia\\
$^3$Centre for Astrophysics \& Supercomputing, Swinburne University of Technology, P.O. Box 218, Hawthorn, VIC 3122, Australia \\
$^4$CSIRO Astronomy \& Space Science, Australia Telescope National Facility, PO Box 76, Epping, NSW 1710, Australia}
\begin{document}
\date{24/08/2012}

\pagerange{\pageref{firstpage}--\pageref{lastpage}} \pubyear{2012}

\maketitle

\label{firstpage}

\begin{abstract}
The Australian Square Kilometer Array Pathfinder (ASKAP) will revolutionise
our knowledge of gas-rich galaxies in the Universe. Here we present
predictions for two proposed extragalactic ASKAP neutral hydrogen (\HI) emission-line 
surveys, based on semi-analytic models applied to cosmological N-body simulations.
\noindent The ASKAP \HI All-Sky Survey, known as WALLABY, is a shallow 3$\pi$ survey 
($z$ = 0 -- 0.26) which will probe the mass and dynamics of over $6\times 
10^{5}$ galaxies. A much deeper small-area \HI survey, called DINGO, aims
to trace the evolution of \HI from $z = 0 - 0.43$, a cosmological volume of $4 \times 10^{7} \Mpc^3$,
detecting potentially $10^{5}$ galaxies. 
\noindent The high-sensitivity 30 antenna ASKAP core (diameter $\sim$2 km) will provide 
an angular resolution of 30 arcsec (at $z=0$). Our simulations show that the 
majority of galaxies detected in WALLABY (87.5\%) will be resolved. About 5000 
galaxies will be well resolved, i.e. more than five beams (2.5 arcmin) across 
the major axis, enabling kinematic studies of their gaseous disks. This number 
would rise to $1.6 \times 10^5$ galaxies if all 36 ASKAP antennas could be
used; the additional six antennas provide baselines up to 6 km, resulting in
an angular resolution of 10 arcsec. For DINGO this increased resolution is 
highly desirable to minimise source confusion; reducing confusion rates from a maximum of 
10\% of sources at the survey edge to 3\%.
\noindent We estimate that the sources detected by WALLABY and DINGO will span four orders of 
magnitude in total halo mass (from $10^{11}$ to $10^{15} \Msol$) and nearly seven orders 
of magnitude in stellar mass (from $10^{5}$ to $10^{12} \Msol$), allowing us to 
investigate the process of galaxy formation across the last four billion years.

\end{abstract}


\section{Introduction}
\label{Introduction}

Neutral hydrogen (\HI) is an ubiquitous tracer of large scale structure in 
the Universe. It allows us to study the physical and dynamical 
processes within galaxies, including the kinematic properties of structures 
such as bars, disks and warps. Each galaxy \HI spectrum provides a large
set of galaxy properties, for example the systemic velocity, the integrated 
flux density and the velocity width. These are used to derive the galaxy 
distance, its gas mass and its total dynamical mass, respectively. The gas
mass is also a good indicator of ongoing star formation. 

The evolution of \HI is of fundamental importance to understanding the 
build-up of both the stellar and gas masses within galaxies as well as the method
by which galaxies accrete their material. Due to 
the inherent signal weakness of the 21-cm hyperfine splitting transition, the
detection of \HI emission in distant galaxies requires high resolution and 
high sensitivity observations.
It is a crucial window into galaxy formation over time; and while we wait for the next generation of large-scale
\HI surveys, we will explore their potential via N-body simulations. 

\begin{table*}
\begin{center}
\begin{tabular}{|c|c|c|c|c|} \hline
Survey & Sky area [$\sqd$] (resolution) & Velocity range [$\kms$]& Detections & Reference \\ \hline
HIPASS$^1$ & 29343 ($15\farcm5$) &  $-1280 < cz < 12700$ & $\sim 5300$ &~\citet{Barnes:01} \\
HIJASS & 1100 ($12'$) &  $-1280 < cz < 12700$ & $222$ &~\citet{hijass} \\
ALFALFA$^2$ & 7000 (4$'$) &  $-1600 < cz < 18000$ & $\sim 15000$&~\citet{Haynes:11} \\
EB\HI S & 21400 (9$'$) & $ cz < 21000$ & Underway &~\citet{Kerp:11} \\
FAST$^3$ & 4000 (3$'$) & $z < 0.4$ & $2\times 10^{6}$&~\citet{Duffy:08a} \\
SKA$^4$ & 20000 (12$''$) &  $z < 1.5$& $10^{9}$ &~\citet{AR} \\
\hline
\end{tabular}
\medskip\\
\caption{We summarise here the key large area \HI surveys that have been completed, or near to completion, as well as 
future projects. $^1$We have included several catalogues; the HIPASS Bright Galaxy Catalogue~\citep{Koribalski:04}, the
southern hemisphere \HI catalogue~\citep{Meyer:04} and its northern
extension ($\delta < 25\degr$;~\citealt{Wong:06}). $^2$For ALFALFA detections we quote the values from the 40\% 
catalogue which at the time of writing was the most complete published results. $^3$The survey considered here consisted of
600s integration times on source, lasting a year. $^4$The SKA as considered in~\citet{AR} would
contain 50\% of the collecting area within 5 km and we therefore use this to determine the typical resolution although in
reality baselines may extend to 3 orders of magnitude larger.}
\label{tab:allsurveys} 
\end{center}
\end{table*}

In Table~\ref{tab:allsurveys} we present a, non-exhaustive, list of large area \HI surveys that have
either been completed, or are currently ongoing, as well as several future surveys of note. 

Several large-scale \HI surveys were obtained with the 64-m Parkes 
telescope, made possible by the innovative 21-cm multibeam system which 
consists of 13 dual-polarisation feed horns and a powerful correlator~\citep{StaveleySmith:96}. 
Most prominent among them is the \HI Parkes All Sky Survey 
(HIPASS;~\citealt{Barnes:01}). We also include the (unfinished) Northern counterpart, the \HI
Jodrell All Sky Survey (HIJASS;~\citealt{hijass}) which utilised a 4-beam receiver.
We list two major surveys currently underway in Table~\ref{tab:allsurveys}.
The first is the Arecibo Legacy Fast ALFA survey (ALFALFA) which utilises the Arecibo L-Band Feed Array (ALFA)
7-beam receiver on the 305m Arecibo dish. 
The second is the Effelsberg-Bonn \HI Survey (EB\HI S;~\citealt{Kerp:11}), carried out 
with the recently installed 7-beam system on the 100m dish.
Finally we present two future surveys in the table, the first is the Chinese-built Five-hundred metre Aperture Spherical 
Telescope (FAST;~\citealt{KARST}) which utilises a 19 beam receiver; and 
Square Kilometre Array (SKA\footnote{SKA homepage: \url{www.skatelescope.org}}).

There are three precursor instruments to the SKA: the Murchison Widefield 
Array (MWA;~\citealt{Lonsdale:09}), the Meer-Karoo Array Telescope (MeerKAT;~\citealt{meerkat}) and the 
Australian SKA Pathfinder (ASKAP;~\citealt{Johnston:08,Deboer:09}). 

Here we will focus on planned \HI surveys with ASKAP; which is currently under construction
in the Murchison Radio Astronomy Observatory in Western Australia. 
ASKAP will consist of 36 antennas (12-m diameter), of these
30 antennas are located within a 2-km diameter circle. ASKAP's large field 
of view - 30 square degrees - provided by novel phased array feeds~\citep{paf}
make ASKAP a 21-cm survey machine. 

In the following sections we introduce two ASKAP \HI surveys: the shallow \HI All-Sky 
Survey (known as WALLABY;~\citealt{WALLABY}), and the deep, 
but small-area \HI survey (known as DINGO;~\citealt{DINGO}). 

\subsection{WALLABY}\label{sec:wallaby}
The {\em Widefield ASKAP L-band Legacy All-sky Blind surveY} 
(WALLABY\footnote{WALLABY homepage: 

\url{www.atnf.csiro.au/research/WALLABY}}) 
is a large project led by B\"arbel Koribalski and Lister Staveley-Smith. 
WALLABY proposes to observe $\sim$75\% of the sky ($-90\degr < \delta <  +30\degr$) 
out to a redshift of $z=0.26$. In the WALLABY 
proposal~\citep{WALLABY} it is estimated that 500,000
galaxies will be detected over the full survey area (assuming an angular 
resolution of 30\arcsec), of these $\sim$1000 galaxies will be spatially 
well resolved (i.e. $>$5\arcmin in angular extent; or $>$10 beams). 

To achieve a much higher angular resolution of 10\arcsec the full ASKAP
configuration (36 antennas with baselines up to 6 km) is needed. Given the high computational 
cost of spectral line imaging of large volumes at such high resolution we are considering to obtain "postage 
stamps". These are high resolution mini-cubes formed at the position and velocity of particularly interesting 
galaxies determined a-priori. This is possible because the 21-cm data are collected by the full array and will
serve many survey science projects.
For simplicity in this work we assume that all galaxies will be imaged using the $6\,\rm km$ 
baselines as we are also interested in probing the issue of 
resolving out objects and hence studying what fraction of galaxies become non-detected. 

The WALLABY goals, outlined in detail in the ASKAP Survey Science Proposal~\citep{WALLABY}, 
are to examine the properties, 
environment and large-scale distribution of gas-rich galaxies.  In summary, 
WALLABY will study galaxy formation and the 
missing satellite problem in the Local Group, evolution and star-formation in 
galaxies, mergers and interactions in galaxies, the \HI mass function and its 
variation with local environment, processes governing the evolution and 
distribution of cool gas at low redshift, and the nature of the cosmic web. 

WALLABY will also be able to investigate cosmological parameters. For example,
we will be able to measure the matter power spectrum in the local Universe.
Furthermore, we should be able to constrain the equation of state of Dark 
Energy to better than 20\%~\citep{Duffy:12b}. 
Other tests that have been proposed for 
such a large \HI survey are studying the coherent bulk flows of galaxies on 
large scales~\citep[e.g.][]{Burkey:04,Abate:08}, 
the measurement of Baryonic Acoustic Oscillations and the Hubble 
constant~\citep{Beutler:11}, the measurement of the rms mass fluctuations, 
$\sigma_8$ \& growth factor, $f$~\citep{Beutler:12} and the surface 
brightness dimming of objects with intrinsic brightness, the so-called Tolman 
test~\citep{Khedekar:11}. 

In this paper we also consider a proposed Northern Hemisphere HI survey that,
when combined with WALLABY, will provide a true \HI all-sky survey. A 
proposal has been submitted to the Astron Westerbork Synthesis Radio Telescope 
facility to carry out the {\em Westerbork Northern Sky
\HI Survey} (WNSHS\footnote{WNSHS homepage: 

\url{http://www.astron.nl/~jozsa/wnshs}}). 
This project is led by Guyla Jozsa, and will target the Northern sky using new phased-array feeds
that can instantaneously observe $8\sqd$. WNSHS as proposed will likely be slightly deeper survey
than WALLABY with a resolution similar to the full $6\,\rm km$ baseline of ASKAP. The science goals
for WNSHS are similar, and due to filling in the sky coverage missed by ASKAP, 
entirely complimentary to WALLABY. As currently designed there are two survey modes for WNSHS,
the first is a limited integration time of 4 hours per observation, which would perfectly match the 
expected sensitivity for the WALLABY survey. The second is a deeper, 12 hour pointing, which would decrease
the flux limits by 1.7 and open up a much deeper exploration (more in line with the deep ASKAP survey, DINGO,
discussed later). We consider both survey modes, with a 4 hour WNSHS model + WALLABY, termed `ALL SKY'
and the deeper WNSHS case on it's own.
In this study we will show that the ALL SKY combination of WALLABY and the 4 (12) hour pointing WNSHS 
could potentially detect of order 800k ($>10^6$) galaxies.

\subsection{DINGO}\label{sec:dingo}
The {\em Deep Investigation of Neutral Gas Origins} 
(DINGO\footnote{DINGO homepage:

\url{http://www.physics.uwa.edu.au/~mmeyer/dingo}})
survey is led by Martin Meyer and consists of deep and ultradeep phases 
which differ in area and depth. In the first phase, the survey 
proposes to target five non-contiguous fields, $150\sqd$ in total, out to $z$ = 0.26. While 
the redshift range is the same as for WALLABY, the integration per field is 500 hours
($>60\times$ longer) providing 8$\times$ better sensitivity. Where feasible the target fields 
will be selected to overlap with the Galaxy And Mass Assembly (GAMA~\citealt{Driver:09}) survey. 
The second phase is proposed to consist of two ultra-deep fields, $60\sqd$ 
in total, over the redshift range $z$ = 0.1 to 0.43. Very long integration
times with 2500 hrs per field will enable DINGO UDEEP to probe 
the evolution of \HI over the last 4 billion years of cosmic time.

DINGO is designed to probe the \HI universe out to the maximum redshift possible,
enabling the evolution in key cosmological parameters (such as the cosmic \HI density)
and the \HI mass function to be measured. As well as this it will sample a sufficient volume to measure the 
two-point correlation function and the halo occupation distribution function as a function of 
redshift. Overlaps with GAMA fields will provide matching stellar properties
for the DINGO \HI detections. 
Alternatively, for the \HI non-detections in these fields, the optical redshifts from 
GAMA enable the possibility of \HI spectral stacking to extend the effective limit of DINGO.

\subsection{ASKAP \HI Predictions}
Estimating the performance of the proposed ASKAP surveys, WALLABY and DINGO, 
is a challenging theoretical problem due to both the large cosmic volumes probed by these surveys ($3.26$ and 
$0.04\, \rm Gpc^3$ respectively) and the low detection mass threshold ($\sim10^{8} \Msol$). 

Previous work on simulating \HI in cosmological
simulations using fully hydrodynamical simulations~\citep[e.g.][]{Popping:09,Altay:11,Duffy:12a} have been limited to 
smaller volumes $(<100 \hMpc)^3$ and hence are unsuitable for making accurate predictions of the large variety
of structure found with ASKAP. Instead we can use $N$-body simulations, which are computationally
cheaper to run and hence can simulate larger regions of the Universe. Creating galaxy properties in simulations that 
do not explicitly track the gas is called semi-analytic modelling. The low computational cost of running a galaxy formation 
model atop an existing $N$-body simulation is such that the various parameters in the model can be tuned to successfully
recreate numerous observational constraints.

Several recent attempts have been made to split the cold gas from the semi-analytic model 
into atomic and molecular hydrogen components~\citep[e.g.][]{Obreschkow:09b,Power:10,Lagos:11}. 
Using observational the constraint, from~\citet{Blitz:06} and~\citet{Leroy:08}, that the molecular - atomic ratio \Htwo / \HI
depends on the 
local Interstellar Medium pressure, ~\citet{Obreschkow:09b} found that the \HI mass function does not strongly
evolve strongly until $z>1$. Using a similar methodology~\citet{Power:10} found that this conclusion holds for several
semi-analytic models.~\citet{Lagos:11} differs slightly in that they calculate the \Htwo / \HI ratio in the semi-analytic model
itself and form stars from the molecular component alone. The non-evolving prediction of the \HI mass function holds as
before, with some interesting new predictions at low \HI masses $<10^{8} \hMsol$ which are below the typical thresholds
for the galaxies found in ASKAP but which the SKA might detect in large numbers.

In this study we seek a simple recipe to guarantee that the \HI masses in the simulation are similar to those observed 
locally. Hence, rather than adopting one of the previously mentioned techniques, we argue for a more transparent method. 
One possibility could be to randomly assign \HI masses based on the locally observed \HI mass 
function~\citep[e.g.][]{hipass,Martin:10} to dark matter haloes in an existing $N$-body simulation. However, there are several 
advantages in using the galaxy properties from semi-analytic modelling. First, confusion estimates based on 
realistic clustering of galaxies become possible. In addition we can probe the effects of modest evolution in the \HI 
abundance of the galaxies as their cold gas mass changes in the simulation. 
Finally, one can examine the galaxy stellar properties from the catalogue with the \HI selected sample.

To these ends we make use of a semi-analytic catalogue of galaxies to form a mock lightcone for ASKAP as detailed in 
Section~\ref{sec:method}, with the method by which we create \HI gas masses for each galaxy in the catalogue given in 
Section~\ref{sec:creatingHI}.
We consider the sensitivity of a radio telescope in Section~\ref{sec:hisignal} and the impact that realistic spatial and spectral 
features for the simulated galaxies will have when observed by instruments such as ASKAP in Section~\ref{sec:resolvedhi} and Section~\ref{sec:resolvespec}. 
In Section~\ref{sec:wallaby} and Section~\ref{sec:dingo} we detail the anticipated number counts for
WALLABY and DINGO respectively, along with the 
ability of ASKAP to probe large distributions of galaxy properties in such a flux limited sample.
The issue of confusion of \HI sources is considered in Section~\ref{sec:confusion} for the particular case of DINGO.
We consider the problems of identifying optical counterparts to the undetected \HI sources to enable the stacking of their
spectra in Section~\ref{sec:misidentification}. We compare the HI surveys in Section~\ref{sec:compare}.
Finally we conclude in 
Section~\ref{sec:conclusion} and emphasis the need for zoom-in, high resolution images around galaxy detections,
so-called `postage-stamps', to limit the effects of confusion on source counts as well as guarantee a large sample of 
resolved galaxies that can be modelled using their velocity field information.

\section{Methodology}
\label{sec:method}

The galaxy catalogue is created using the semi-analytic model (SAM) of~\citet{Croton:06} 
to produce galaxies based on the underlying dark-matter-only {\sc Millennium Simulation}~\citep{Springel2005}. 
This sample of galaxies accurately 
recreates the observed stellar mass function with a combination of Supernovae feedback and, crucially to this model,
feedback at the high mass end from active galactic nuclei. The cosmology used in the {\em Millennium Simulation}
is $(\Omega_M = 0.25, \, \Omega_\Lambda = 0.75, \, \Omega_b = 0.045, \, \sigma_8 = 0.9, \, h = 0.73)$. This differs 
from the current best fitting cosmological model based on measurements of Supernovae and the Cosmic Microwave Background~\citep{Komatsu:11},
most significantly in the relatively high $\sigma_8$ value adopted in the simulation. This will have the effect of creating 
haloes that, at a fixed total mass of $10^{11} \, (10^{14}) \hMsol$, are $\sim 30\% \, (20\%)$ too concentrated~\citep{Duffy:08a}.
We ignore the effects of baryons on the Dark Matter halo, the baryonic backreaction, 
which typically increase the concentration but, with sufficient feedback (e.g. by an accreting Supermassive Blackhole),
gas can be expelled to actually reduce the concentration relative to the Dark Matter only simulation~\citep[e.g][]{Duffy:10}.
At all masses the haloes will be $\sim 5\%$ too spherical but with identical spin distributions~\citep{Maccio:08} relative to the `true' 
cosmology.

We then utilised the {\it Theoretical Astrophysical Observatory (TAO)}\footnote{\url{http://tao.it.swin.edu.au/mock-galaxy-factory}}, which is 
a {\sc cloud}-based web application which produces mock catalogues from different cosmological simulations
and galaxy models in the form of a lightcone (detailed in~Bernyk et al. {\it in preparation}) to create an all sky galaxy catalogue 
extending to $z=0.26$ for the WALLABY survey and to $z=0.43$ for the narrower but deeper DINGO survey. Although the DINGO surveys
are not contiguous fields (they are placed around the sky to lessen cosmic variance) we make the simulated lightcone a single field
of the same total area, this will not have any impact on the expected galaxy counts.
This tool creates a lightcone
through the simulation volume, stitching a randomly rotated cube to the far edge to minimise the repeated structure in a cone. Objects that
are broken up by this rotation of the cube are separately reattached to preserve the 1 and 2 -halo terms. The closest snapshot in redshift 
from the full simulation catalogue is used where possible to accurately trace the evolution of galaxies. This resource provides stellar, dark 
matter and cold gas masses (amongst many other quantities) for all galaxies within the lightcone.

We use the flux limit calculations from~\citet{Duffy:12b} and figures for ASKAP given in Table~\ref{tab:telescope_values} for several HI
surveys. In Table~\ref{tab:survey_values} (and ~\ref{tab:survey_results}) we provide survey parameters and summarise the results of our simulations. The respective surveys are WALLABY (as described in Section~\ref{sec:wallaby}), All-Sky (WALLABY plus a Northern hemisphere
extension with the same resolution and sensitivity, essentially the 4hour WNSHS survey), WNSHS (the 12hour deep survey with 
1.7$\sigma$ better sensitivity than WALLABY) and finally DINGO DEEP and UDEEP (as given in Section~\ref{sec:dingo}). 

\begin{table}
\begin{center}
\begin{tabular}{|c|c|c|} \hline
Parameter & ASKAP core & ASKAP full\\ \hline
$N_{\rm dish}$ & 30 & 36 \\
$\aeff$ (${\rm m^{2}}$) & 2669 & 3211 \\
$\tsys$ ($K$) & 50 & 50 \\
Max Baseline (km) & 2 & 6 \\
Ang Res (z=0) & 30'' & 10'' \\
Field-Of-View ($\sqd$) & 30 & 30 \\
\hline
\end{tabular}
\medskip\\
\caption{We summarise here the strawman values of ASKAP~\citep{Johnston:08} as it is likely to be operational
for the first several years of its spectral-line use, namely a reduced baseline model `core' which utilises the inner 30 dishes across 
a maximum $2\, \rm km$ baseline rather than `full' with 36 dishes, and $6\,\rm km$ maximum baseline, 
which will be used for higher-resolution observations. Ultimately, one may be able to use the full 36 dishes of ASKAP using short
baseline pairs only such that the $30''$ resolution is attained for all dishes. In this case the number of pairs is only marginally increased 
however and therefore our sensitivity calculation with 30 dishes is still valid; for simplicity we assume only 30 dishes are used for the 
$30''$ observation. Furthermore, it is not clear what weighting scheme will be employed by ASKAP for the high resolution, $6\, \rm km$,
baselines and hence whether the full 20\% flux sensitivity will be realised (for the case of natural weighting). To that end we 
conservatively assume that both the low and high resolution modes have the same flux levels. 
The effective area assumes an aperture efficiency of 80\% and assumes the use of cross-correlations only (see text). Uniquely with
ASKAP the Field-of-View is independent of frequency, and hence across the depth of the survey, with increasing Field-of-View as 
the observed wavelength lengthens balanced by modifying the number of beams formed on the sky.}
\label{tab:telescope_values} 
\end{center}
\end{table}

\begin{table*}
\begin{center}
\begin{tabular}{|c|c|clclc|c|} \hline
\multirow{2}{*}{Parameter} &  \multirow{2}{*}{${\rm WALLABY}$}& \multirow{2}{*}{ALL SKY}&   \multirow{2}{*}{${\rm WNSHS}$} &   \multirow{2}{*}{${\rm DINGO ~DEEP}$} & ${\rm DINGO ~UDEEP}$  \\ 
 & & & & & $\Omega_{\rm \HI}$ Fixed, \, \, {\bf R Fixed} \\ \hline
{Sky Obs [${\rm deg^{2}}$]} & 30940 & {41252} & 10313 & 150 & \,\,\,\,\,\,\,\,\,\,\,\,\,\,\,\, 60  \\
{Survey  [hrs]} & 9600 & 14800 & 16900 & 2500 &  \,\,\,\,\,\,\,\,\,\,\,\,\,\,\,\, 5000 \\
{Redshift range} & 0 - 0.26 & 0 - 0.26 & 0 - 0.26 & 0 - 0.26 &  \,\,\,\,\,\,\,\,\,\,\,\,\,\,\,\, 0.1 - 0.43  \\ \hline
1$\sigma$ noise [mJy $\kms$] & & & & & \\ 
$\Delta v$ = 3.86$\kms$ & 1.592 & 1.592 & 0.937 & 0.201 &  \,\,\,\,\,\,\,\,\,\,\,\,\,\,\,\, 0.090 \\
$\Delta \nu$ = 100 kHz & 0.681 & 0.681 & 0.401 &  0.086 &  \,\,\,\,\,\,\,\,\,\,\,\,\,\,\,\, 0.039 \\ \hline
{$\rm D_{\rm \HI}$ [kpc]} & $14\pm^{7}_{3}$ & $14\pm^{7}_{3}$  & $14\pm^{7}_{3}$ & $14\pm^{7}_{3}$ & $15\pm^{8}_{4}$, ${\bf 15\pm^{9}_{4}}$  \\ 
{Detected $\rm D_{\rm \HI}$ [kpc]} & $42\pm^{19}_{14}$ & ${42\pm^{19}_{16}}$ & $42\pm^{19}_{16}$ & $41\pm^{15}_{15}$ & $41\pm^{16}_{13}$, ${\bf 42\pm^{17}_{13}}$  \\ \hline

Low Res  & & & & \\
$> 1$ beam $[N_{\rm gal}\,(\%)]$& $542706 \,(87.5)$ & ${726346 \,(87.5)}$ & 327454 (62.4) & $4926$ (9.0) & $759$ (1.4), ${\bf 855 \, (1.4)}$  \\ 
$> 3$ beams $[N_{\rm gal}\,(\%)]$& $30859\,(5.0)$ & ${41397\,(5.0)}$ & 11557 (2.2) & $152$ (0.3) & $0$ (0), ${\bf 0 \, (0)}$  \\ 
$> 5$ beams $[N_{\rm gal}\,(\%)]$& $6205\,(1.0)$ & ${8394\,(1.0)}$ & 2416 (0.5)& $38$ (0.07) & $0$ (0), ${\bf 0 \, (0)}$  \\ 
$> 10$ beams $[N_{\rm gal}\,(\%)]$ & $699\,(0.1)$ & ${946\,(0.1)}$ & 271 (0.05)& $2$ (0.004)& $0$ (0), ${\bf 0 \, (0)}$  \\ 
$> 20$ beams $[N_{\rm gal}\,(\%)]$ & $74\,(0.01)$ & ${109\,(0.01)}$ & 37 (0.01)& $0$ (0) & $0$ (0), ${\bf 0 \, (0)}$  \\ \hline

High Res  & & & & \\
$> 1$ beam  $[N_{\rm gal}\,(\%)]$& $247038\,(100)$ & ${323156 \,(100)}$ & 356447 (100) & $44904$ (94.8) & $37018$ (75.3), ${\bf 41490\, (76.5)}$  \\ 
$> 3$ beams  $[N_{\rm gal}\,(\%)]$ & $239653\,(97.0)$ & ${310109\,(96.0)}$ & 153328 (43.0) & $4926$ (10.4) & $759$ (1.5), ${\bf 855\, (1.6)}$  \\ 
$> 5$ beams  $[N_{\rm gal}\,(\%)]$ & $146280\,(59.2)$ & ${183811\,(56.9)}$ & 31759 (8.9) & $955$ (2.0) & $47$ (0.10), ${\bf 37 \,(0.07)}$  \\ 
$> 10$ beams  $[N_{\rm gal}\,(\%)]$ & $22149\,(9.0)$ & ${26310\,(8.1)}$ & 3761 (1.1) & $115$ (0.2) & $0$ (0), ${\bf 0 \,(0)}$  \\ 
$> 20$ beams $[N_{\rm gal}\,(\%)]$ & $2476\,(1.0)$ & ${2868\,(0.9)}$ & 432 (0.1) & $10$ (0.02) & $0$ (0), ${\bf 0 \,(0)}$  \\ 								
\hline
\end{tabular}
\medskip\\
\caption{We summarise here the proposed survey specific values for the two aspects of DINGO (DEEP and Ultra-DEEP), WALLABY and WNSHS as described in the text. 
For Ultra-DEEP we consider two evolution models; a varying \HI-cold gas fraction such that $\Omega_{\rm HI}$ remains constant and a fixed \HI-cold gas fraction that
results in a modest increase in $\Omega_{\rm HI}$ (the latter given in bold face).
For reference we include the $1\sigma$ rms noise levels in each survey for two bandwidth measures of $3.86\kms$ and 100 kHz. Note 
that with the final decision on the baseline weighting still to be decided, the high resolution mode is conservatively set to the same flux
limits as the low resolution case (therefore the number of resolved systems is a lower limit). 
We list the median and quartile limits of the \HI diameter of the galaxies in each survey, along with the value for those 
galaxies {\it detected} (in the low resolution survey only). We consider what fraction of these systems are 
resolved, i.e. have an angular diameter greater than the angular resolution of the telescope (the low res value is $30''$).
These are then separated into the expected number of galaxies that will be resolved by more than 1, 3, 5, 10 or 20 beams
(in parenthesis are the percentage of detected galaxies in each case). We then consider a postage-stamp reimaging 
around each of these detections at higher resolution; using $10''$ resolution available with the full 6$\rm km$ baseline of ASKAP. 
For the case of WNSHS the highest resolution imaging is $13''$. 
This extra resolution makes a significant difference to the final number counts of well resolved systems and argues for a high resolution
reimaging for strong galaxy detections. The survey number counts are presented in Table~\ref{tab:survey_results}.}
\label{tab:survey_values} 
\end{center}
\end{table*}

\begin{table*}
\begin{center}
\begin{tabular}{|cc|c|clclc|c|} \hline
{\multirow{2}{*}{S/N}} & {\multirow{2}{*}{Distribution}} &  \multirow{2}{*}{${\rm WALLABY}$} &  \multirow{2}{*}{ALL SKY} &   \multirow{2}{*}{${\rm WNSHS}$} &   \multirow{2}{*}{${\rm DINGO ~DEEP}$} & ${\rm DINGO ~UDEEP}$  \\ 
 & & & & & & $\Omega_{\rm \HI}$ Fixed, \, \, {\bf R Fixed} \\ \hline
\multirow{2}{*}{5} 	& $n_{\rm gal}$  & 620123 (738121) & {829843 (988145)}& 524921 (583976) & 54697 (55679) & 53896 (54544), {\bf 59408 (60092)} \\
				    	& $\bar{z}$ & 0.054 (0.059) 	 & {0.054 (0.059) } 	  & 0.071 (0.075) 	 & 0.130 (0.131)   & 0.217 (0.218), {\bf 0.220 (0.221)}  \\ 
\\
\multirow{2}{*}{6} 	& $n_{\rm gal}$ & 462923 (558084) & {619174 (746740)}& 397288 (450584) & 43614 (44690) & 41126 (41720), {\bf 45446 (46145)} \\
					& $\bar{z}$ & 0.049 (0.053) 	& {0.049 (0.053) } 	 & 0.065 (0.069) 	& 0.123 (0.125)   & 0.208 (0.209), {\bf 0.211 (0.212)}  \\ 
\\
\multirow{2}{*}{7} 	& $n_{\rm gal}$ & 359394 (455654) & {480476 (609595)}& 312469 (361622) & 35600 (36652) & 32514 (33111), {\bf 36079 (36763)} \\
					& $\bar{z}$ & 0.045 (0.050) 	& {0.045 (0.050) } 	 & 0.060 (0.065) 	& 0.118 (0.119)   & 0.201 (0.202), {\bf 0.203 (0.204)}  \\ 
\\
\multirow{2}{*}{8} 	& $n_{\rm gal}$ & 287390 (376405) & {384325 (503236)}& 253222 (298782)  & 29693 (30663) & 26556 (27104), {\bf 29337 (29955)} \\
					& $\bar{z}$ & 0.041 (0.047) 	& {0.041 (0.047) } 	 & 0.056 (0.061)  	 & 0.112 (0.114)   & 0.194 (0.196), {\bf 0.196 (0.198)}  \\ 
\\
\multirow{2}{*}{9} 	& $n_{\rm gal}$  & 235502 (317191) & {314858 (424281)}& 209555 (252324)  & 25386 (26377) & 21838 (22389), {\bf 24411 (25000)} \\
					& $\bar{z}$  & 0.038 (0.045) 	 & {0.038 (0.045) }      & 0.052 (0.058)    	  & 0.107 (0.109)   & 0.189 (0.190), {\bf 0.191 (0.192)}  \\ 
\\
\multirow{2}{*}{10} 	& $n_{\rm gal}$ & 196119 (272341) & {262328 (364399)}& 176959 (216630)  & 21980 (22933) & 18407 (18934), {\bf 20654 (21183)} \\
					& $\bar{z}$ & 0.036 (0.043) 	& {0.036 (0.043) }      & 0.049 (0.055)  	 & 0.103 (0.105) & 0.184 (0.186), {\bf 0.186 (0.187)}  \\
\\
\multirow{2}{*}{15} 	& $n_{\rm gal}$ & 94767 (151174) & {127064 (202187)}& 89993 (120789)  & 12402 (13235) & 9027 (9443), {\bf 10142 (10658)} \\
					& $\bar{z}$ & 0.028 (0.035)     & {0.029 (0.035) }      & 0.039 (0.045)       & 0.087 (0.090) & 0.167 (0.170), {\bf 0.167 (0.170)}  \\
\\					
\multirow{2}{*}{20} 	& $n_{\rm gal}$ & 54688 (99285) & {73349 (132934)}& 54558 (79653) & 8230 (8977) & 4997 (5422), {\bf 5868 (6282)} \\
					& $\bar{z}$ & 0.023 (0.031)  & {0.023 (0.031) }    & 0.032 (0.040)    & 0.077 (0.080) & 0.157 (0.160), {\bf 0.156 (0.159)}  \\	
\\					
\multirow{2}{*}{25} 	& $n_{\rm gal}$ & 35031 (71613) & {47015 (95879)}& 36585 (57474) & 5941 (6620) & 3090 (3416), {\bf 3611 (3981)} \\
					& $\bar{z}$ & 0.020 (0.028)  & {0.020 (0.028) } & 0.028 (0.035)     & 0.069 (0.073) & 0.150 (0.152), {\bf 0.148 (0.152)}  \\	
\\										
\multirow{2}{*}{30} 	& $n_{\rm gal}$ & 23808 (54753) & {32070 (73334)}& 25874 (44165)  & 4576 (5188) & 2049 (2334), {\bf 2427 (2713)} \\
					& $\bar{z}$ & 0.018 (0.025)  & {0.018 (0.025) }  & 0.025 (0.032)    & 0.063 (0.068) & 0.144 (0.147), {\bf 0.143 (0.146)}  \\									
\hline
\end{tabular}
\medskip\\
\caption{We summarise here the proposed survey specific values for the two aspects of DINGO (DEEP and Ultra-DEEP) and WALLABY~\citep{WALLABY}
as well as a Northern extension plus WALLABY (`ALL SKY') and the WNSHS survey, similar to Table~\ref{tab:survey_values}.
We present two numbers for the predicted galaxy counts, and their mean redshift, reflecting
the effects of including the reduction of signal-to-noise by spatially resolved galaxies, as
demonstrated in Fig.~\ref{fig:detection_res}. The brackets ignore this effect and therefore have a larger 
galaxy number count. There are second numbers in boldface for DINGO UDEEP which is our evolutionary model.
The \HI mass function is allowed to evolve by retaining the same \HI - cold gas mass conversion for $z=0$ but applied to
the high redshift cold gas mass function, which increases with redshift in agreement with~\citet{Power:10}. 
There is more cold gas at high redshift, and hence more \HI, which results in the ultra-deep survey outperforming the 
wider angle, lower redshift DINGO survey in terms of number of galaxies detected.}
\label{tab:survey_results} 
\end{center}
\end{table*}

\subsection{Estimating \HI in Galaxies}
\label{sec:creatingHI}

\begin{figure}
  \begin{center}
   \epsfig{figure=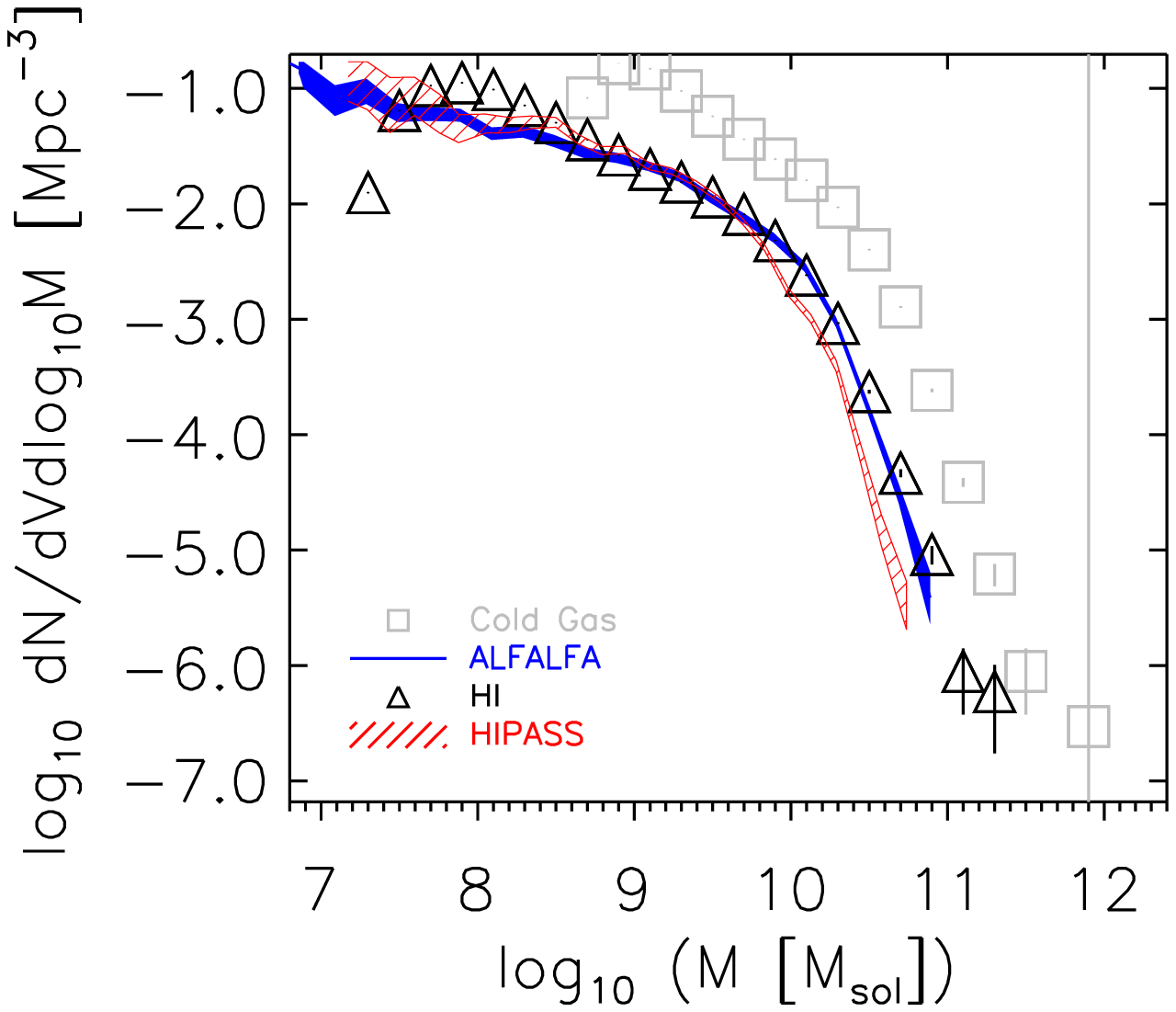, scale=0.5} 
    \caption[mass functions]
            {Here we present several mass functions to indicate the methodology of obtaining a `realistic' observed \HI mass per
            halo. We start with the cold gas mass function from the~\citet{Croton:06} semi-analytic model (grey square) and compare with
            the observed \HI mass function from ALFALFA (or equivalently HIPASS) in the blue (red) solid (hatched) line
            (\citealt{Martin:10,hipass}, respectively). We then apply a cold gas - \HI conversion in the form
            of a broken power law function, given in Eqn.~\ref{eqn:minimisefn}, before recalculating the mass function and once again
            comparing to the ALFALFA observations. This is repeated using an AMOEBA code to minimise the difference, resulting in the 	
            black triangles which are our final \HI mass function. 
            Note that we adopted a cold gas mass cut of $10^{8.5} \, {\rm M}_{\odot}$ as the resolution limit,
            in accordance with~\citet{Power:10}, which is why the grey squares abruptly end at low mass. 
            The error bars are Poissonian (with the highest mass bins containing only two or three objects).}
    \label{fig:himassfn_orig}
  \end{center}
\end{figure}

The~\citet{Croton:06} model produces a list of cold gas masses for the galaxies that we can attempt to break into neutral atomic, molecular hydrogen 
and ionised hydrogen components located near to the galaxy.
At the high mass end we would envisage significant fractions of neutral hydrogen to be in molecular form~\citep{Leroy:08,Saintonge:11},
with the low mass end predominantly atomic hydrogen with a component of ionised hydrogen due to the ionising cosmic
UV background~\citep[e.g.][]{Quinn:96}. The cold gas - \HI mass fraction has been previously studied in semi-analytic works as both a constant, 
e.g.~\citet{Power:10} found $0.54$ after considering He abundances and typical ionisation fractions of the gas,
as well as a function of the halo properties themselves~\citep[e.g.][]{Obreschkow:09a}. As we are
interested in creating a realistic mock catalogue rather than understanding the \HI properties of galaxies we 
apply a reduction ratio $R = M_{\rm \HI} / M_{\rm cold}$ to convert cold gas to \HI so as to recover the observed \HI mass function.\footnote{In this work we take the ALFALFA mass function~\citep{Martin:10} to model the galaxies on but note that this typically results in more HI rich
galaxies than if we used HIPASS~\citep{hipass}. The latter mass function results in 16\% fewer galaxies, for example, in the DINGO DEEP survey.} 
We find that a functional form suggested by~\citet{Yang:03} for creating a conditional luminosity function by
assigning stellar mass to dark matter haloes is particularly suitable for this conversion
\begin{equation}
\label{eqn:minimisefn}
R = \left( \frac{M_{\rm \HI}}{M_{\rm cold}} \right)_{0} \left[ \left( \frac{M_{\rm cold}}{M_{\star}} \right)^{-\alpha} +  \left( \frac{M_{\rm cold}}{M_{\star}} \right)^{\beta}  \right]^{-1} \,,
\end{equation}
where the fraction of \HI in the cold gas is $\left( M_{\rm \HI} / M_{\rm cold} \right)_{0} = 0.41$ at a characteristic cold gas mass $M_{\star} = 8.8 \times 10^{10} \Msol$
with faint end slope $\alpha = 0.52$ and bright end slope $\beta = 0.56$ for a sample of galaxies in the redshift range $0.02 < z < 0.07$ centred on the mean 
redshift of the WALLABY survey $\bar{z} \approx 0.05$ (as calculated in~\citealt{Duffy:12a}). 
We refer to the survey predictions using this conversion scheme as 
$\emph{Fixed R}$. Interestingly, the normalisation of the fraction for the low redshift sample is similar to the value suggested by~\citet{Power:10}. 

As~\citet{Power:10} demonstrated the cosmic density of cold gas within semi-analytic models typically increased by $0.2$ dex over the redshift range $z=0-0.43$
and hence we can expect that our predictions with $\emph{Fixed R}$ will demonstrate mild evolution in the resultant HI mass function. However, we can also refit
the mass function in several redshift slices such that we can recover a non-evolving $\Omega_{\rm \HI}$; we have calculated these, fairly arbitrary, conversions for the
higher redshift sample in DINGO UDEEP. We call this conversion the \emph{Fixed $\Omega_{\rm \HI}$} model. 
We include the different predictions for the \emph{Fixed} \emph{R} / \emph{$\Omega_{\rm \HI}$} cases in Table~\ref{tab:survey_values}.
However, in all plots and discussions in the text we will consider only the later case as it is both the most conservative in terms of galaxies detected as well as in better 
agreement with the data, which typically shows little evolution in cosmic \HI density to $z=3$~\citep[e.g.][]{Rao:06,Lah:07,Noterdaeme:09,Prochaska:09,Meiring:11}.
As shown in Fig.~\ref{fig:himassfn_orig} this method creates an \HI mass function in excellent agreement with the data. 

\begin{figure}
  \begin{center}
   \epsfig{figure=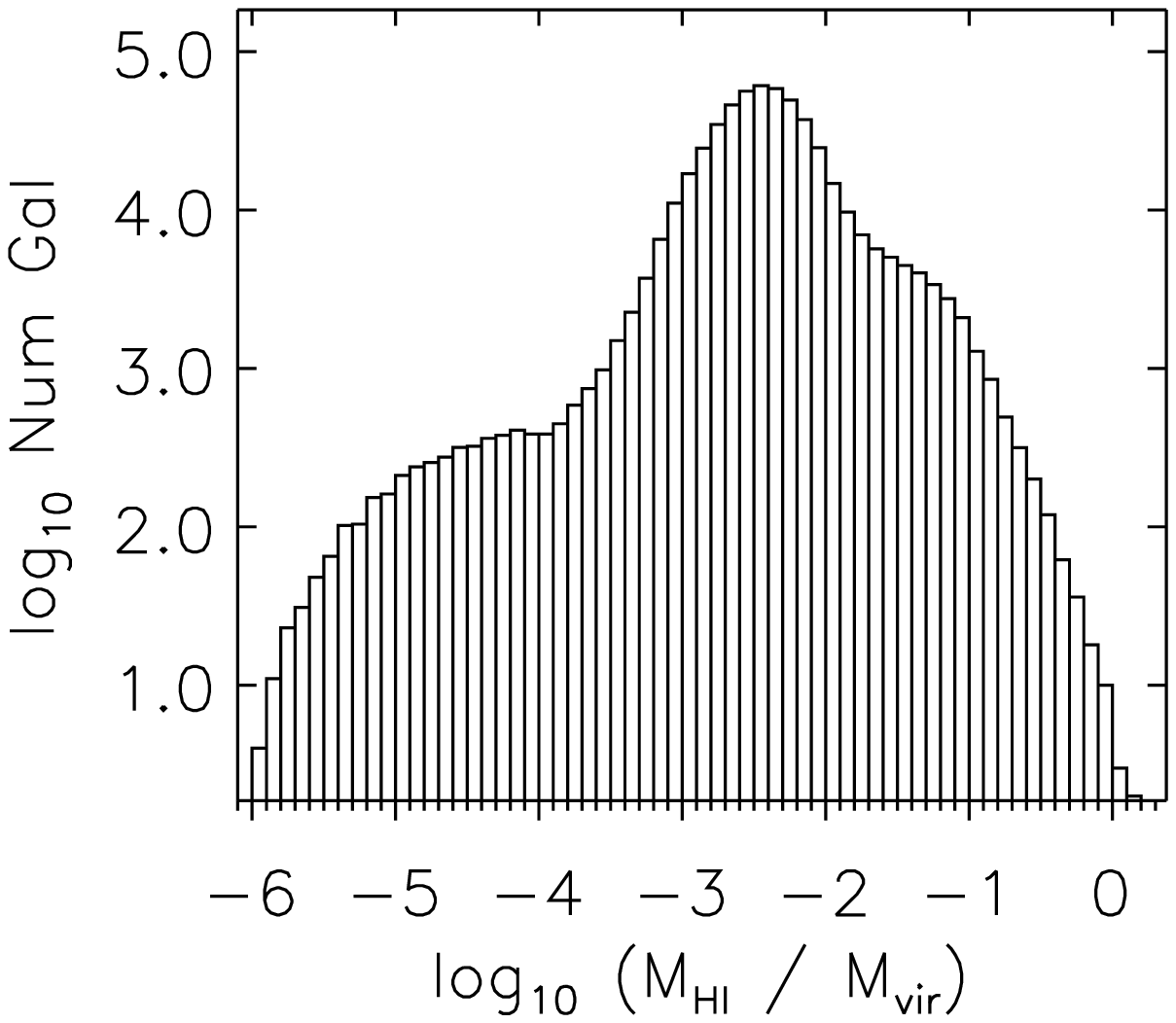, scale=0.5}
   \epsfig{figure=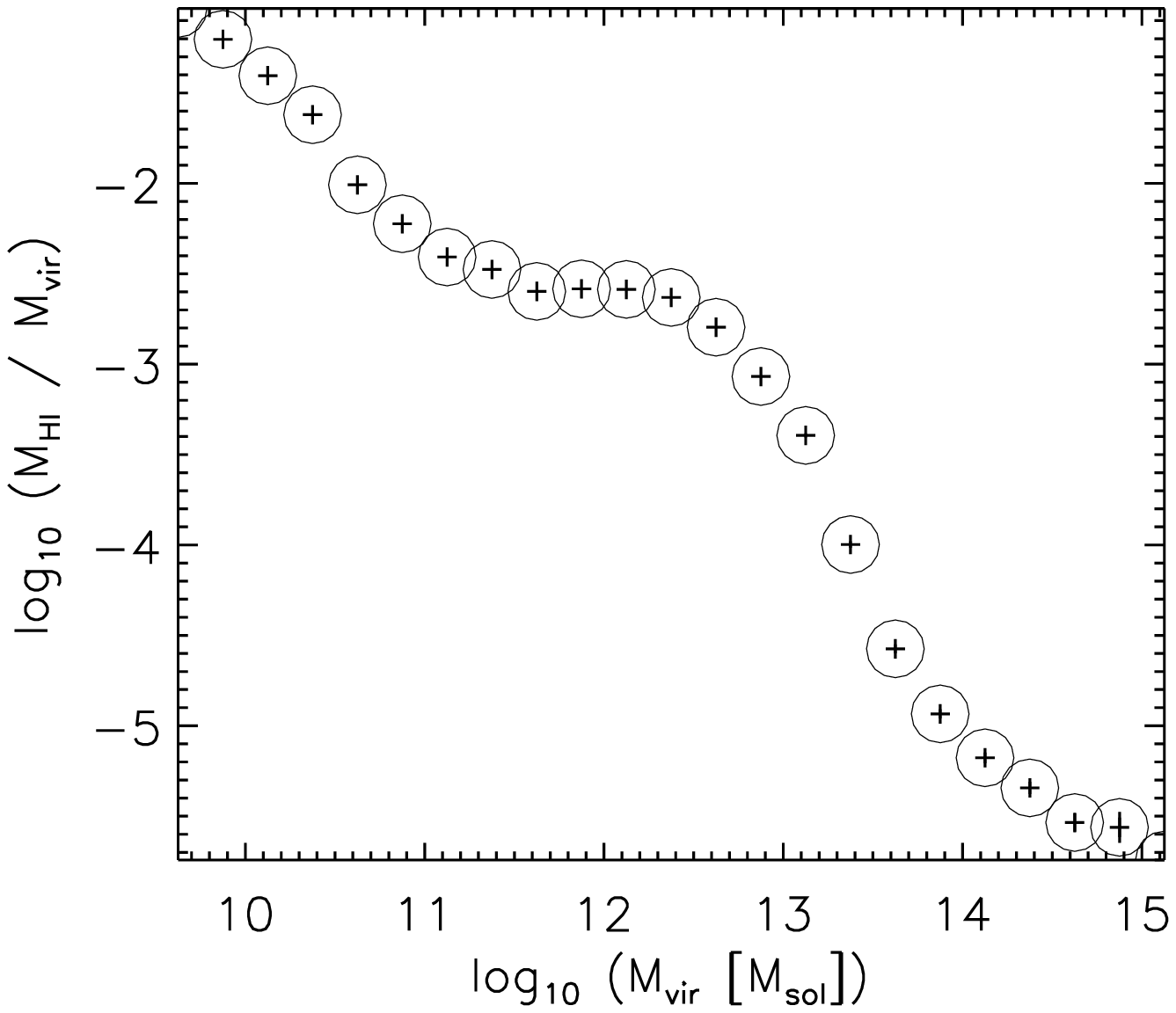, scale=0.5} 
    \caption[\HI halo ratio]
            {In the top panel we show a histogram of the \HI mass - Halo mass ratio of all systems in the DINGO DEEP volume, indicating that our 
            methodology results in a non-linear
            relation between \HI mass and underlying halo mass. This is not a surprise as the SAM catalogues have a cold gas mass that reflects a 
            star formation and merger history that will all act to create a complex interplay between the gas phase and the underlying halo mass. In the bottom panel
            we plot the ratio of \HI - halo mass as a function of halo mass and find that typically more massive systems are relatively \HI deficient. We have estimated
            errors within the points by bootstrapping the haloes but the 68\% confidence limits are within the point size for all but the most massive datapoint.}
    \label{fig:HI_MVir_ratio}
  \end{center}
\end{figure}

This simple, yet effective, method produces a complex \HI - halo mass distribution that we examine in Fig.~\ref{fig:HI_MVir_ratio}. 
In the top panel we consider the histogram distribution of the \HI - halo mass ratio, finding that the majority of systems have
an \HI mass that is approximately 1\% of the total halo mass. This distribution sharply drops off to higher ratios with a strongly asymmetric
tail to lower ratios which extends to systems incredibly \HI poor. If we consider which systems are, relatively, the most devoid of \HI
we find (from the bottom panel of Fig.~\ref{fig:HI_MVir_ratio}) that they are the most massive systems in the simulation, 
in agreement with observations~\citep[e.g.][]{Kilborn:09}.

\subsection{Resolution Limits of the Simulation}
For such low \HI detection limits as proposed in the DINGO and WALLABY surveys there may be a potential issue with the resolution 
limits of the semi-analytic catalogue itself. This is because we have conservatively set the minimum \HI mass of the catalogue to be
the same as the cold gas mass limit, i.e. $10^{8.5} \, {\rm M}_{\odot}$. 
We can try and estimate the number of sources we may mass by using
an observed HI mass function and extrapolating below this mass limit. To that end we use the empirical technique presented 
in~\citet{Duffy:12b}, which include realistic rotation widths and HI disk sizes for the galaxies, to determine what fraction of galaxies would 
have been detected below the limiting HI mass cut for the proposed ASKAP surveys presented here. 

Using empirical relations from~\citet{Duffy:12b} we find that the DINGO DEEP survey would detect 64859 galaxies and of these 2871 
have an angular extent greater than $30\arcsec$. When we impose a limiting \HI mass of $10^{8.5} \, {\rm M}_{\odot}$ we find 60314 
galaxies are detected, of these 2785 are larger in angular extent than $30\arcsec$. This means that by constraining the semi-analytic 
catalogue to a limiting \HI mass of $10^{8.5} \, {\rm M}_{\odot}$ we underestimate the DINGO DEEP survey by approximately 4500 
galaxies, or nearly 8\%. Of these missed objects the majority are unresolved, with only 1\% having an angular extent greater than 
$30\arcsec$.

For the low redshift WALLABY survey the full catalogue with no mass cut is estimated to be 672660 galaxies and 471129 of these are 
resolved by the $2\rm km$ baseline mode of ASKAP. When a limiting \HI mass of $10^{8.5} \, {\rm M}_{\odot}$ is assumed, the galaxy 
count drop to 631409. Of these objects, 462495 (522981) are resolved by an angular beam of $30\arcsec$. 
In other words we are potentially underestimating the WALLABY catalogue by 40k galaxies, or 7.5\% similar to the case of DINGO DEEP, but nearly a quarter of these galaxies would be marginally resolved.

The predictions for the high redshift DINGO UDEEP catalogue are essentially unchanged by setting the limiting \HI mass of 
$10^{8.5} \, {\rm M}_{\odot}$. Therefore we caution the reader that the catalogue created in this work will likely underpredict the lower 
redshift DINGO DEEP and WALLABY surveys by 8\%, with the vast majority of these systems being unresolved point sources for the 
case of DINGO but as many as a quarter of this missing population marginally resolved in the WALLABY survey. With these caveats in 
mind we now consider the manner in which we can use the semi-analytic catalogue to create a realistic HI survey.

\subsection{Creating \HI surveys}\label{sec:hisignal}
The monochromatic luminosity, $L_{\nu}$ from a cloud of \HI (of mass $M_{\rm \HI}$) with a 21-cm line emission line profile $\phi(\nu)$ (a narrow function of unit area, where the frequency $\nu$ is
close to the rest frequency, $\nu_{12} = 1.42040575 \rm \, GHz$) is
\begin{equation}
L_{\nu} = \frac{3}{4} h \nu_{12} A_{12} \frac{M_{\rm \HI}}{m_{\rm H}} \phi(\nu)\,,
\end{equation}
where $A_{12}$ is the Einstein coefficient for the \HI spin-flip transition. We can relate this to the total integrated flux of the source in terms of 
the bolometric luminosity $S_{\rm TOT} = L_{\rm TOT}/(4\pi d^2_{\rm l}(z))$ (Eqn 3.89 from~\citet{peacockbook}) at a luminosity distance 
$d_{\rm l} = D (1+z)$, where $D$ is the comoving distance and $L_{\rm TOT}=\int L_{\nu} d\nu$. 
The total flux is the flux density, $S_{\nu}$, integrated over frequency:
\begin{equation}
S_{\rm TOT} = \int S_{\nu} d\nu = \frac{3}{16\pi} \frac{h \nu_{12} A_{12}}{d^2_{\rm l}(z)} \frac{M_{\rm \HI}}{m_{\rm H}} \int \phi(\nu) d\nu \,.
\end{equation}
This equation simplifies to 
\begin{equation}\label{eqn:mass limit}
 \frac{S_{\rm TOT}}{\rm Jy\, \rm Hz} \approx  \frac{M_{\rm \HI}}{\rm M_{\odot}} \frac{1}{49.8}\left(\frac{d_{\rm l}(z)}{\rm Mpc}\right)^{-2} \,.
\end{equation}
In the case of a boxcar line profile, $S_{\rm TOT}=S_{\nu}\Delta\nu$ where $\Delta\nu$ is in the observed frame
of the galaxy. Note that a given rest-frame frequency or velocity width is measured by the observer to be $(1+z)$ {\it narrower} 
(i.e. a galaxy will span fewer observed channels at higher redshifts for the same intrinsic rotation width). This becomes important
when calculating the signal-to-noise of the observation in the following section.

We emphasise that this 
formula implicitly assumes an optically thin approximation for the \HI i.e. that there is negligible self-absorption of the \HI flux, in the galaxies themselves.
This is a valid assumption for massive late-type systems with mean correction factors of $19\%$ for Sb, $16\%$ for Sbc \& Sc and $4\%$
for Sa and Sab galaxies found in~\citet{Zwaan:97} after averaging over angle of inclinations from correction factors found in~\citet{Haynes:84}. In~\citet{Haynes:84}
smaller dwarf systems had a larger correction factor, however, the majority of the WALLABY and DINGO detections lie above these mass limits. 
Therefore to a good approximation we can ignore the effects of inclination on the detectability of the galaxy in terms of total flux received from the galaxy. 
There are observational consequences of the inclination of the galaxy in terms of the observed velocity width and the resolving out of extended sources which we consider 
in Section~\ref{sec:resolveout} that are fully consistent with this optically thin assumption.

\subsubsection{Detection Limits}
As detailed in~\citet{Duffy:08a} and references therein, the expected thermal noise for a dual 
polarisation single beam, single dish telescope is given by
\begin{eqnarray}\label{eq:flux noise} 
\sigma_{\rm noise}=\sqrt{2}\frac{k\tsys}{A}\frac{1}{\sqrt{\Delta t \, \Delta \nu}}\end{eqnarray}
for an observing time of $\Delta t$ within a bandwidth $\Delta \nu$, which is assumed to be the ASKAP frequency resolution of $18.31 \, \rm kHz$ (a velocity width $\Delta V = 3.86\, \kms$ at $z=0$), 
where $k=1380\,{\rm Jy}\,{\rm m}^2\,{\rm K}^{-1}$ is the Boltzmann constant and $T_{\rm sys}$ is the system temperature. 

\citet{Thompson:99} showed that for a dual polarisation interferometer, the noise is reduced by a further $\sqrt{2}$ if the area $A$ is taken to
be that of a single element, i.e.:
\begin{eqnarray}\label{eq:interferometer} 
\sigma_{\rm noise}=\frac{k\tsys}{A}\frac{1}{\sqrt{\Delta t \, \Delta \nu}},\end{eqnarray}
so that for a full interferometer with $N(N-1)/2$ baseline permutations:
\begin{eqnarray}\label{eq:array noise} 
\sigma_{\rm noise}=\sqrt{2} \frac{k\tsys}{A}\frac{1}{\sqrt{N(N-1) \Delta t \, \Delta \nu}}.\end{eqnarray}
The effective area of an 
ASKAP dish is the geometric area of a $12\,\rm m$ diameter dish, $a$, reduced by the aperture 
efficiency, expected to be $\alpha_{\rm eff} \approx 0.8$~\citep{Johnston:08}. Therefore, Eqn.~\ref{eq:array noise} can be re-written:
\begin{eqnarray}\label{eq:array noise final} 
\sigma_{\rm noise}=\sqrt{2} \frac{k\tsys}{A_{\rm eff}}\frac{1}{\sqrt{\Delta t \, \Delta \nu}},\end{eqnarray}
where $A_{\rm eff}=\alpha_{\rm eff}a\sqrt{N(N-1)}$. We assume $N=30$ dishes within the 2-km core since, due to processing limitations, the full ASKAP array of 36 dishes is unlikely to be initially available. Note the 
well-known similarity between Eqn.~\ref{eq:flux noise} and \ref{eq:array noise final} for large $N$.

The flux density limit for an observation, $S_{\rm lim}$, depends on the required signal-to-noise ratio $(S/N)$:
\begin{eqnarray}\label{eq:fluxlimit_channel} S_{\rm lim}= (S/N) \sigma_{\rm noise}\,.\end{eqnarray}
A galaxy is less easily detected if its flux is spread across a number of frequency
channels, resulting in lower flux density. For a boxcar profile of rest-frame velocity width $W$, a line profile will be spread over $N_{\rm ch} = (W / \Delta V) / (1+z)$ channels, where $\Delta V$ is the $z=0$ ASKAP velocity resolution of $3.86\kms$. 
As the noise is assumed to be Gaussian the uncertainty in our measurement of the mean flux density 
is reduced by the square root of the number of independent channels, or samples. Hence, for detection:
\begin{eqnarray}\label{eq:fluxlimit_first} S_{\rm TOT} \, {\rm (Jy\, Hz)} > (S/N) \frac{\sigma_{\rm noise}}{\sqrt{N_{\rm ch}}}N_{\rm ch}\, \Delta \nu \,.\end{eqnarray}
We can thus observe a galaxy if it has integrated flux greater than the integrated noise with a signal-to-noise cut off
\begin{eqnarray}\label{eq:fluxlimit} S_{\rm TOT} \, {\rm (Jy\,Hz)} > (S/N) \sigma_{\rm noise}\Delta \nu \sqrt{N_{\rm ch}}\,,\end{eqnarray}
where we note again that the frequency bandwidth $\Delta \nu$ is fixed by the ASKAP correlator to be $18.31 \, \rm kHz$ and that the number of channels
an observed object spans will decrease as $(1+z)$ due to the redshifting of the \HI line, aside from any additional evolutionary effects.

\begin{figure}
  \begin{center}
   \epsfig{figure=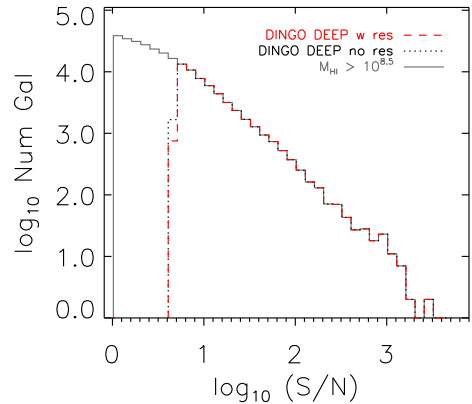, scale=0.5} 
    \caption[detection curve]
            {We consider the number of galaxies distributed as a function of signal to noise for the 
            DINGO DEEP survey. 
            The \HI masses of the galaxies are 
            converted from the cold gas mass from the~\citet{Croton:06} semi-analytic model as given by Eqn.~\ref{eqn:minimisefn} and 
            converted into an \HI signal to noise ratio by applying Eqn.~\ref{eqn:mass limit} and \ref{eq:fluxlimit}. The grey solid histogram is the 
            initial source population
            for all galaxies with cold gas masses above $10^{8.5} \, {\rm M}_{\odot}$ (found by~\citet{Power:10} to be the resolution limit of the 
            Millennium Simulation).
            We then calculate the observed galaxy counts for a DINGO survey in black dotted, with a special 
            case in red dash for galaxies that are
            spatially resolved which suffer a loss of signal as described in Sec~\ref{sec:resolveout}. 
            Galaxies that are larger or face-on will be more resolved,
            although the impact on numbers for a deep survey like DINGO is marginal. Several objects lose nearly an order of magnitude in 
            signal-to-noise
            as a result of this effect which may impact kinematic studies of these objects which demand more robust data than a simple
            detection.}
    \label{fig:detection_res}
  \end{center}
\end{figure}

\subsubsection{Disk diameter}\label{sec:resolvedhi}
As considered in~\citet{Duffy:12b} a potential issue when using interferometers 
with high angular resolution is the resolving out of galaxies more extended than the beam. 
With at least $2 \, \rm km$ baselines assumed for the initial survey phase of ASKAP we will certainly have to consider extended faint sources. 
We now summarise the procedure used in~\citet{Duffy:12b} to determine this issue. 
Although we have accurate \HI galaxy masses in the SAM catalogue we do not have a well-defined disk size
and therefore utilise an empirically derived relation between this mass and the observed \HI diameter, $D_{\rm \HI}$ (defined to be the region inside 
which the \HI surface density is greater than $1M_{\odot}\,{\rm pc}^{-2}$). From~\citet{BR,VS} we have 
\begin{eqnarray}\label{eq:size mass relation} \frac{D_{\rm \HI}}{\rm kpc}=\left( \frac{M_{\rm \HI}}{M_{\rm norm}} \right)^{\gamma}\,,
\end{eqnarray}
where they find an index of $\gamma = 0.55$ and a normalisation mass of $M_{\rm norm} = 10^{6.8} \, \rm M_{\odot}$. 
We then convert this diameter to an on-sky 
angular scale using the angular diameter distance $d_{\rm A}(z)$. Note that this relation is of critical importance 
to our results and we are confident that this relationship is well supported as more recent observations by~\citet{Noordermeer:05} have found 
{\it identical} best fit results to~\citet{BR,VS} for an entirely new galaxy sample, observed with a different 
telescope. Although these works are limited to higher column densities than might be probed by the 
deeper HI survey DINGO we believe that it is still appropriate to use Eqn.~\ref{eq:size mass relation}
as the majority of the HI signal will be from the high column density material described by this relation.

The distribution of the typical diameters of the galaxies found in the various ASKAP \HI surveys after selection effects are taken into account are shown in 
Table~\ref{tab:survey_values} as well as the fraction of the galaxies in each survey which are resolved by ASKAP. 

To test the sensitivity of our
results to this relation we have refit the~\citet{Noordermeer:05} results for the case of a uniform average surface density of $1M_{\odot}\,{\rm pc}^{-2}$, i.e. 
$\gamma =0.5$, and find a normalising mass of $M_{\rm norm} = 10^{6.5} \, \rm M_{\odot}$ with only $10\%$ greater reduced $\chi^{2}$ than the
case where $\gamma$ is unconstrained. In using this more theoretically motivated diameter-mass relation our overall number counts, after considering 
resolving of extended sources, decrease by less than 1\% in the case of DINGO, typically the detected galaxies are of order 1kpc larger
than the case where $\gamma $ is free to vary.
We are therefore confident that this critical assumption of a fixed \HI diameter - mass relation is both supported by the data and that small variations around this
best fit solution make percent-level modifications to the numbers quoted in Table~\ref{tab:survey_values} and hence can be ignored. 
Evolution in the \HI diameter-mass relation as modelled in~\citet{Obreschkow:09c} was found to be negligible.

\subsubsection{Telescope resolution}\label{sec:resolveout}
The angular resolution of ASKAP will be initially
limited to $30''$ at 21cm wavelength using the central $2\,\rm km$ core. 
Typically, for radio telescopes, the beam area increases like $\lambda^{2}\propto (1+z)^2$. However, this is not the case for the ASKAP phased array feeds 
which have a roughly fixed covering area as a function of redshift.

The effect of resolution on the detectability of galaxies depends on the efficiency and nature of galaxy-finding algorithms. Development 
of several algorithms is underway for ASKAP and preliminary testing has been undertaken by \citet{Popping:12}. For the present 
purposes, we will make the assumption that the ASKAP source finders will be able to combine neighbouring independent pixels of a 
spatially-resolved galaxy to improve signal-to-noise ratio in the same manner as for a spectrally-resolved galaxy (see Eqn.~\ref{eq:fluxlimit_first}). We therefore assume that we have optimal spatial smoothing of our sources in the sky plane.

The number of pixels the galaxy spans is approximated by the ratio of the galaxy on-sky area, $A_{\rm gal}$, and the 
beam area $A_{\rm beam}$, each of which is calculated below. In practice the beam is convolved with the galaxy, 
meaning that even when a galaxy is just unresolved by the telescope (i.e. covers one pixel) we would still expect to lose signal.
This convolution is represented by an additional
factor of unity added to the number of beams, reducing the signal-to-noise by $\sqrt{2}$ in this matched case. 
Therefore the total flux of the galaxy, i.e. the left hand side of Eqn.~\ref{eq:fluxlimit}, is reduced by the geometric factor $\sqrt{1+ A_{\rm gal} / A_{\rm beam}}$
(an example of this is shown in Fig.~\ref{fig:detection_res}).

To calculate the galaxy area we approximate the true 3D galaxy profile by an elliptical cylinder with a spectral profile that is independent of position. We further assume a random inclination to the observer, $\theta$, where the case of $\theta = 0$ corresponds to face-on and 
$\theta=\pi/2$ to edge-on. The projected area of the galaxy is $\pi (D_{\rm \HI}/2)^2 (B/A)$ where A and B are the major and minor axes 
respectively the ratio of which, B/A, is equal to $\cos(\theta)$, 
although in calculating this we limit the smallest axis ratio for spirals to 0.12 in accordance with~\citet{Masters:03}. 

For the natural Gaussian antenna distributions described 
in~\citet{Staveley-Smith:06} and modelled in~\citet{Gupta:08}, the Full-Width Half Maximum beam extent for ASKAP is 
$\Omega_{\rm FWHM} \approx 1.4 \lambda / 2000 \, \rm m$. The beam area, $A_{\rm beam}$, is therefore 
given by $\pi \Omega_{\rm FWHM} / (4 \ln 2)$, which we compare to the area of the galaxy (which has been
multiplied by the square of the cosmological angular diameter distance, i.e. we have included $(1+z)^4$ dimming).

\begin{figure}
  \begin{center}
  \epsfig{figure=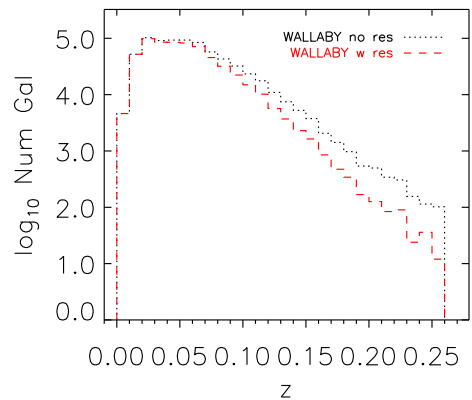, scale=0.5} \\
    \caption[detection curve]
            { We calculate the observed galaxy counts for a WALLABY survey. 
            The black dotted histogram represents the detections possible if all the 
            flux of a galaxy was recovered, while the red dashed histogram is
            the situation if extended galaxies are resolved out. This last case is described in Sec~\ref{sec:resolveout}. 
            Larger galaxies, or those face-on, will be more resolved by more synthesised beams and hence lose signal; with WALLABY 
            losing nearly 15\% of galaxies through this effect.}
    \label{fig:galcount_wallaby}
  \end{center}
\end{figure}

For the current design of ASKAP, with a $2\, \rm km$ baseline, this results in $12.5\%$ of galaxies in WALLABY 
being resolved out. This effect is particularly an issue at the high-z end of the survey where galaxies are faint 
(see Fig.~\ref{fig:galcount_wallaby}). Overall, despite being more spatially resolved, face-on galaxies are 
easier to detect in the ASKAP surveys considered here. This is shown in Fig.~\ref{fig:costheta_res}, where a 
small increase of face-on systems are recovered from the DINGO survey relative to the input uniform cosine 
distribution.

We note that a natural antenna weighting would result in a 20\% increase in flux sensitivity for the high resolution case of ASKAP, 
with the additional six dishes being used. However it is still undecided which weighting scheme will be employed and hence we 
conservatively assume the same flux limit for both the high and low resolution ASKAP images.

\subsubsection{Spectral resolution}\label{sec:resolvespec}
We calculate the velocity width, $W$, of the galaxy from the intrinsic linewidth width, $W_{\rm e}$,  for a given angle of 
inclination $\theta$. The intrinsic linewidth of a galaxy, corrected for broadening, has been shown empirically to be related to the \HI mass 
by~\citep{BriggsRao:1993, hijass}
\begin{eqnarray}
\label{eq:velocity-mass relation}
\frac{W_{\rm e}}{420\, \kms}=\left( \frac{M_{\rm \HI}}{10^{10}M_{\odot}} \right)^{0.3}\,, \end{eqnarray} 
although we note that this relation shows a large dispersion, especially for dwarf galaxies. We have tested the consequences
of this scatter by including an observed $0.2$ dex lognormal dispersion in the velocities and find that 2\% more galaxies are detected
(we henceforth ignore this negligible factor but note that for Tully-Fisher studies it will have to be modelled).
The linewidth of a galaxy, $W_\theta$, which subtends an angle $\theta$ between 
its spin axis and the line-of-sight can be computed using the Tully-Fouque rotation 
scheme~\citep{TFq}
\begin{eqnarray}\label{eq:TFq}
\lefteqn{({W_{\rm e} \sin(\theta)})^{2} = W_\theta^{2} + V_{\rm o}^{2}  - 
2{W_\theta}{W}\left( 1- e^{- \left(\frac{{W_\theta}}{V_{\rm c}}\right)^{2}} \right)} \nonumber \\
& & - 2V_{\rm o}^{2}e^{- \left(\frac{{W_\theta}}{V_{\rm c}}\right)^{2}}\, ,
\end{eqnarray}
where $V_{\rm c} = 120\, \kms$ represents an intermediate transition between the small 
galaxies with Gaussian \HI profiles in which the velocity contributions add quadratically and giant 
galaxies with a `boxy' profile reproduced by the linear addition of the velocity terms. 
$V_{\rm o}\approx 20\, \kms$ is the velocity width due to random motions in the disk~\citep{Rhee96,VS}. 

In cases where $W_\theta>>V_{\rm c}$, one can see that 
$W_\theta=V_{\rm o}+W_{\rm e}\sin\theta$. For $\theta=0$, one finds that 
$W_\theta=V_{\rm o}$, in other words the \HI dispersion in the disk, whereas for 
$\theta=\pi/2$ we recover $W_\theta=V_{\rm o}+W_{\rm e}$ as expected.

\begin{figure}
  \begin{center}
   \epsfig{figure=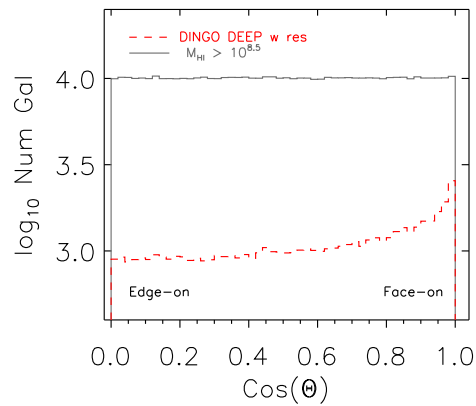, scale=0.5} 
    \caption[angle of inclination detection]
            {We demonstrate in the grey solid histogram the input galaxy angle of inclinations $\theta$ in the DINGO DEEP 
            survey, with an assumed random, uniform distribution in
            cosine $\theta$. 
            After applying the resolution effects and flux sensitivity for this survey we see that there is a preference for galaxies which are
            face-on so that the signal is distributed over the fewest velocity channels, even though this angular extent increases the loss of signal 
            through incoherent addition of the galaxy flux in the plane of the sky by a realistic source finder. The exact mean (median) $\theta$ value 
            in degrees 
            on input for this catalogue is $57.3^{\circ}$ ($60.0^{\circ}$) and after observation the preference for face-on systems is clear as 
            $\theta$ becomes $52.8^{\circ}$ ($54.9^{\circ}$).}
    \label{fig:costheta_res}
  \end{center}
\end{figure}

The small additional broadening effect on the \HI profile due to the frequency 
resolution of the instrument~\citep{bot} is negligible for ASKAP, and is ignored.
Therefore for each galaxy we can uniquely assign an observed linewidth and angular extent on the sky for a random angle of inclination and its \HI mass (see Fig.~\ref{fig:costheta_res}). 

\section{WALLABY}
With the large Field-of-View of ASKAP it becomes feasible to scan the entire sky, in 8 hour integrations, within several years, resulting in 
significant numbers of galaxies detected, $\sim 0.6$ million as given in Table~\ref{tab:survey_results}. 

The effects of resolution on galaxy numbers, as shown in the high res case in Table~\ref{tab:survey_results}, causes a sharp drop in the 
detection rates. Nearly 0.3 million small, faint detections would now drop below the signal-to-noise threshold of the survey due to the 
resolving out of structure. 
As shown in Fig.~\ref{fig:galcount_wallaby}, galaxies are lost throughout the entire redshift range of the survey, including marginally 
resolved systems at high redshift which now drop out of the survey entirely.

The creation of high resolution images of low resolution detections should clearly be limited to only the brightest sources, the exact threshold for which size and signal-to-noise detection should be rescanned is however entirely dependent on the performance of the source finder.

Furthermore source confusion, where multiple \HI sources overlap, is almost never an issue for WALLABY (we direct the reader to Section~\ref{sec:confusion} where this technique is studied for the context of DINGO) in good agreement with our analytic estimates 
in~\citet{Duffy:12b}. Additionally we find for WALLABY detections we can uniquely identify optical counterparts from a given optical
photometric (spectroscopic) redshift catalogue to better than 97\% (99\%) as considered in Section~\ref{sec:misidentification} for the
case of DINGO.

\begin{figure}
  \begin{center}
    \epsfig{figure=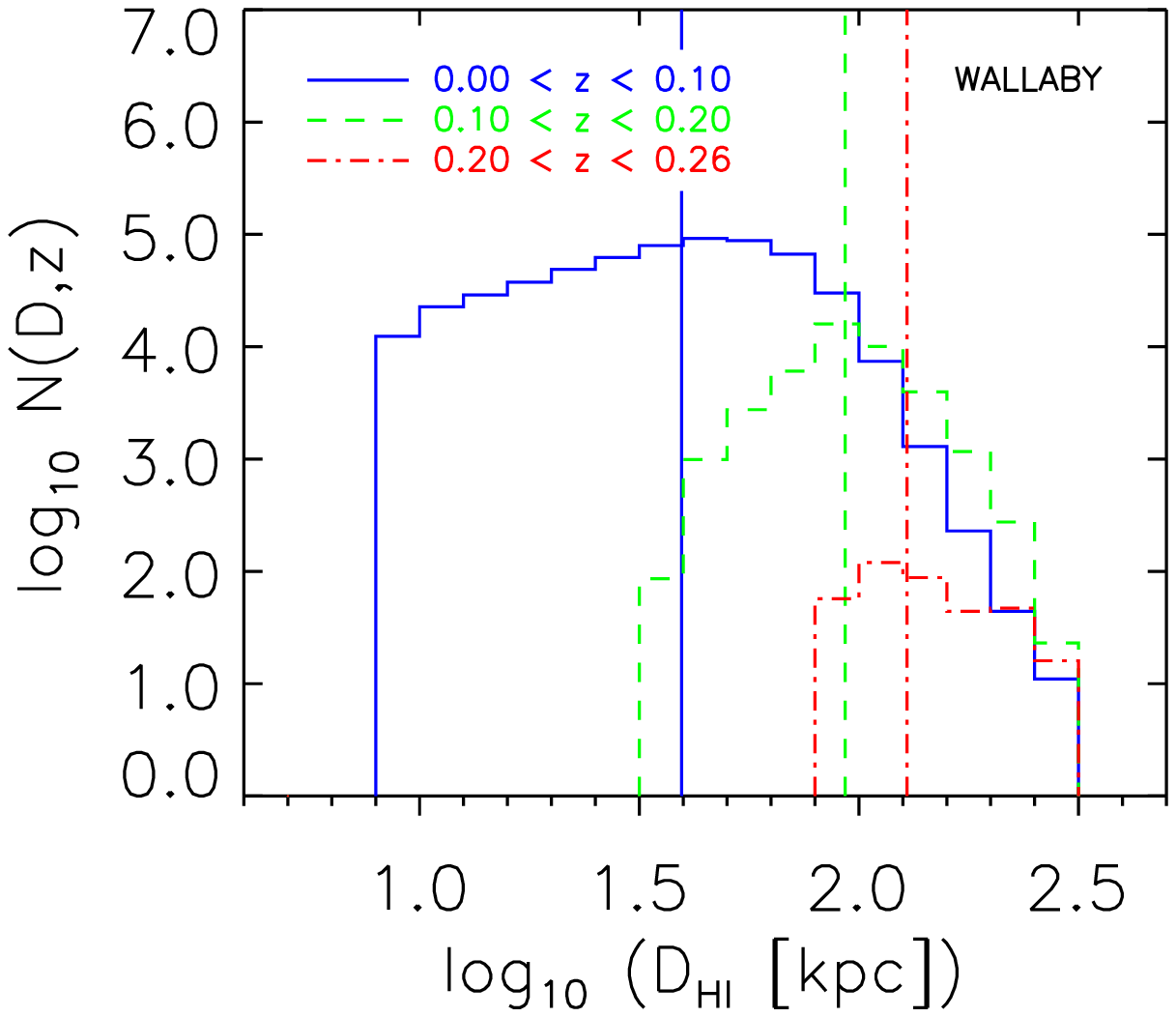, scale=0.45} \\
    \epsfig{figure=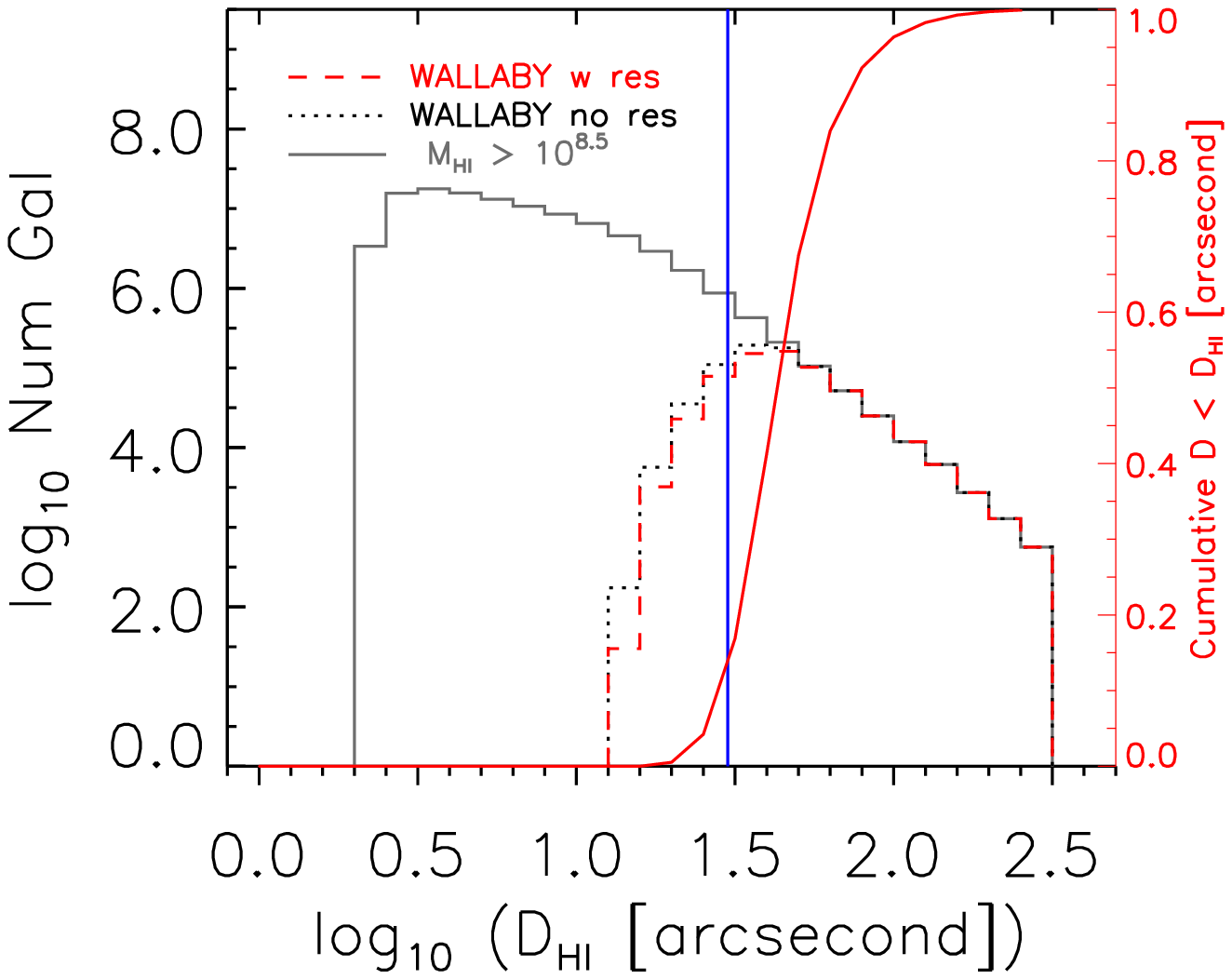, scale=0.45} \\
    \epsfig{figure=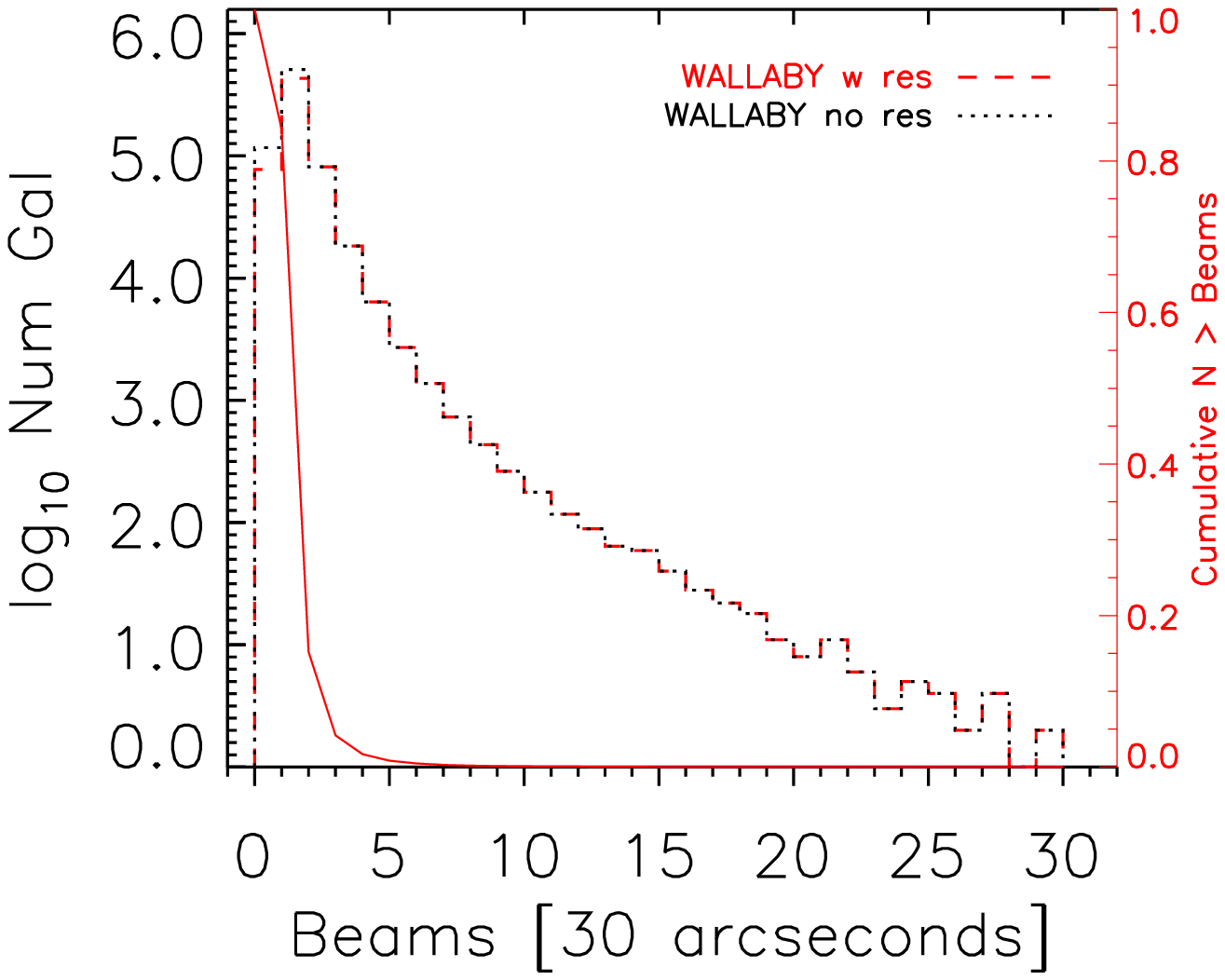, scale=0.45} \\
    \caption[Galaxy Resolved]
            {We consider the sizes of the galaxies as a function of intrinsic physical diameter as estimated using Eqn.~\ref{eq:size mass relation} for
            several redshift cuts (top),
            angular extent on the sky (middle) and the number of beams that the galaxies will be resolved by using ASKAP (bottom) for the
            WALLABY survey.}
    \label{fig:dhi_wallaby}
  \end{center}
\end{figure}

Although the proposed WALLABY survey extends to $z=0.26$ the majority of galaxies detected are at much lower redshifts ($<\bar{z}> \sim 0.05$) and hence will have large angular extents on the sky, which can be expected to be well resolved by the 2km baselines of ASKAP. The distribution of the detections as a function of proper
physical diameter, for different redshift slices, is considered in the top panel of Fig.~\ref{fig:dhi_wallaby}, with the on sky angular extents
of these systems in the middle panel and finally the number of ASKAP beams that these systems are resolved by in the bottom panel.

There are two interesting results from the top panel, the first is that the galaxies detected in a flux limited survey 
are progressively larger in diameter (the solid vertical lines indicate the mean size of the sample) as this scales 
with \HI mass and hence flux. The second is that the WALLABY survey will detect orders of magnitude more 
galaxies within z=0.1 than are detected in the redshift range beyond. 

In the middle panel we consider the angular extent of all simulated galaxies in the WALLABY survey volume (in 
grey solid histogram) and the detected subset with and without resolution effects in black dotted and red dashed respectively. This
subset is typically several arcseconds in extent. The cumulative detection 
given by the red curve shows that at least 50\% of detections are larger than $30\arcsec$. The blue vertical line 
indicates the resolution at 21cm of the 2km baseline of ASKAP; the intersection of this line with the cumulative 
curve indicates that that the majority of ASKAP detections are resolved.

In the bottom panel of Fig.~\ref{fig:dhi_wallaby} we bin the galaxies by the number of ASKAP synthesised beams 
spanning the object. Although most (87.5\%) galaxies are only `marginally' resolved with one or two beams there 
are at least 5\% of all detections resolved by 3 beams (as given by the red curve) and 1\% of 
detections resolved with 5 beams. A non-negligible number of sources are extremely well resolved with
over 30 beams across the galaxy.

The lightcone distribution of these sources, in RA and redshift, are given in Fig.~\ref{fig:lightcone}, with the cosmic web clearly
visible across the volume surveyed, indicating
that a wide range of environments, from clusters to voids, will be probed by the survey. This will in turn allow the measurement of the 
the \HI mass function as a function of environment, as well as various galaxy formation properties such as the stripping of star-forming gas
within the hot haloes of clusters~\citep[e.g.][]{Kilborn:09}.

\begin{figure}
  \begin{center}
    \epsfig{figure=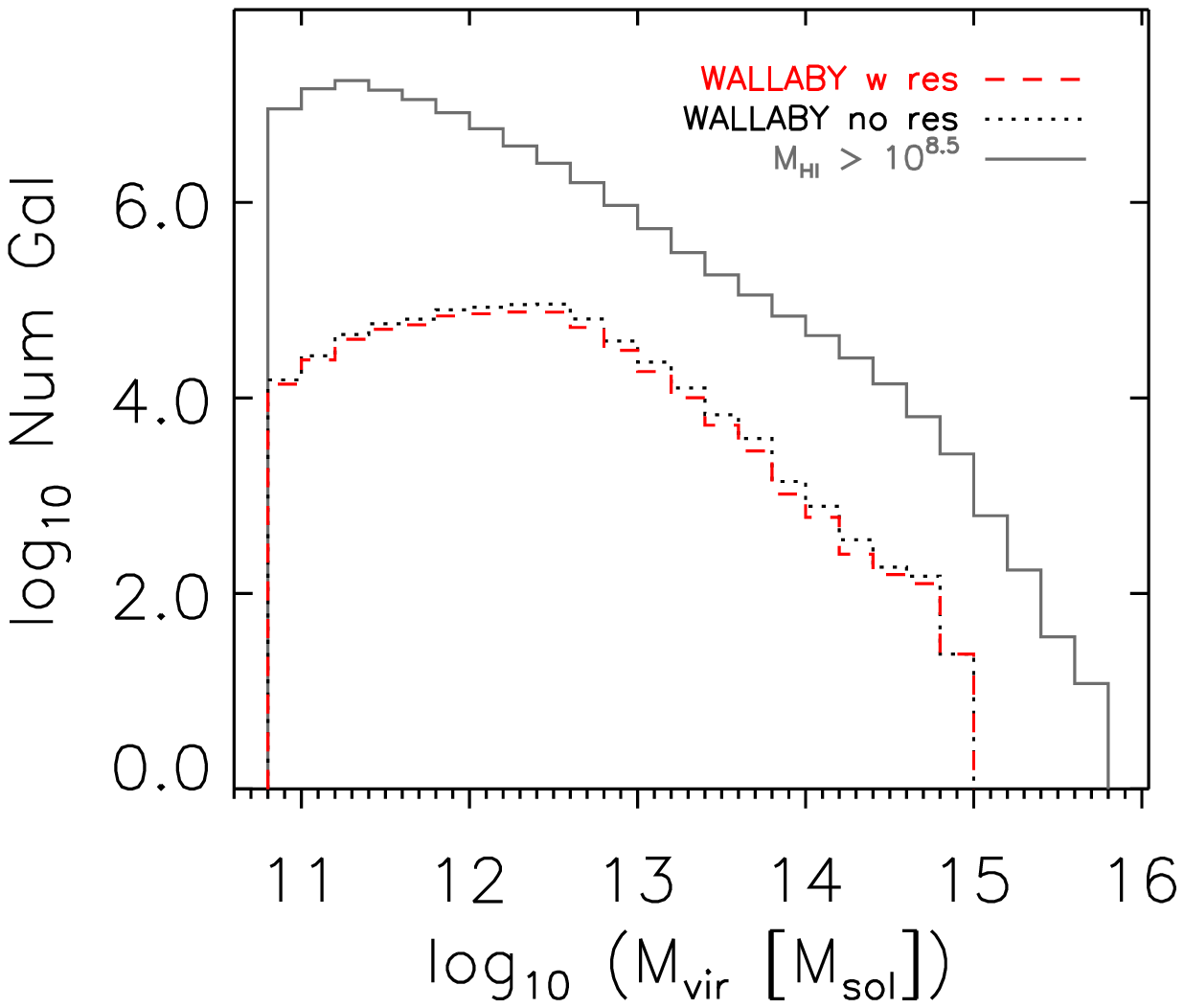, scale=0.45} \\
    \epsfig{figure=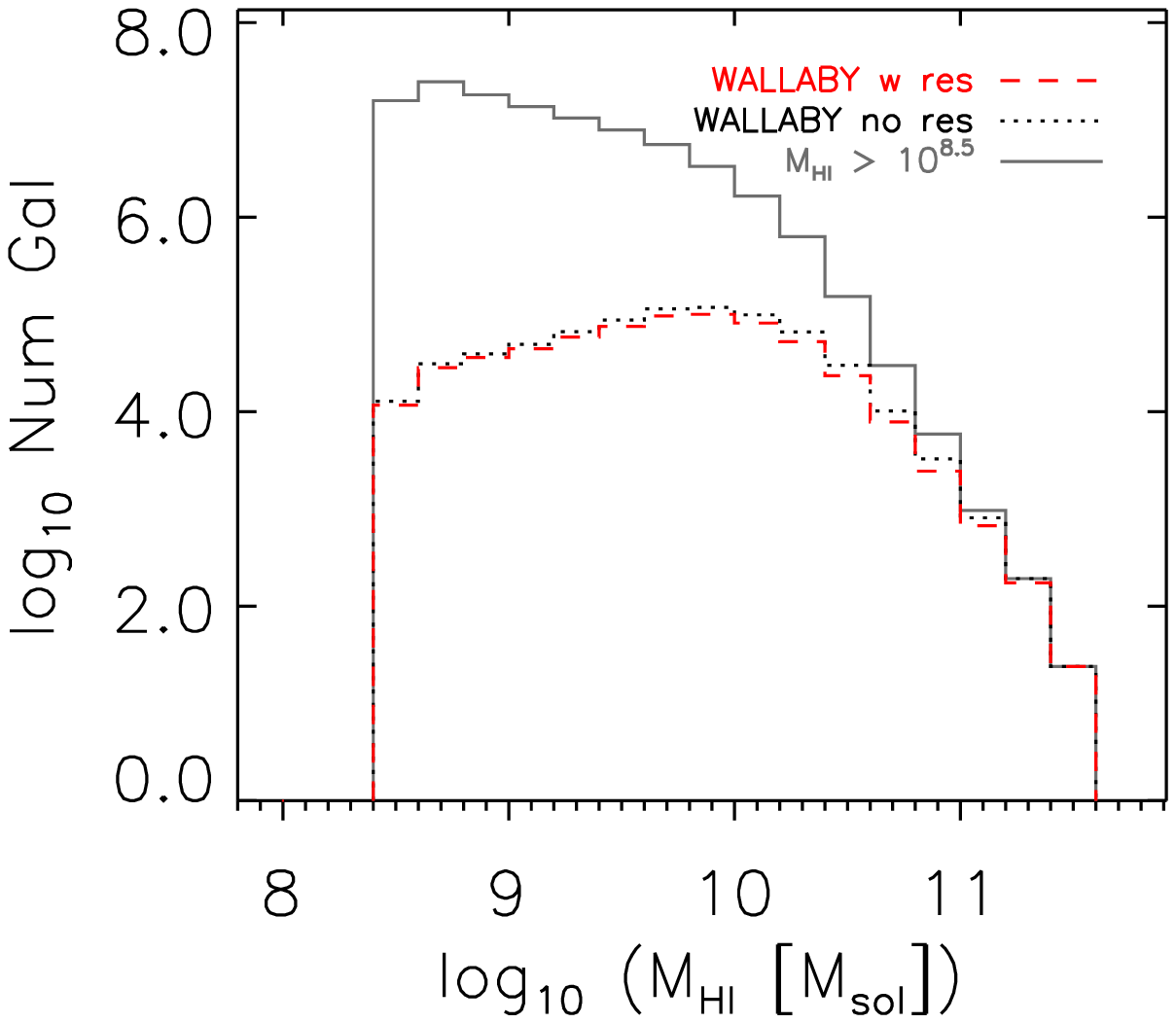, scale=0.45} \\
    \epsfig{figure=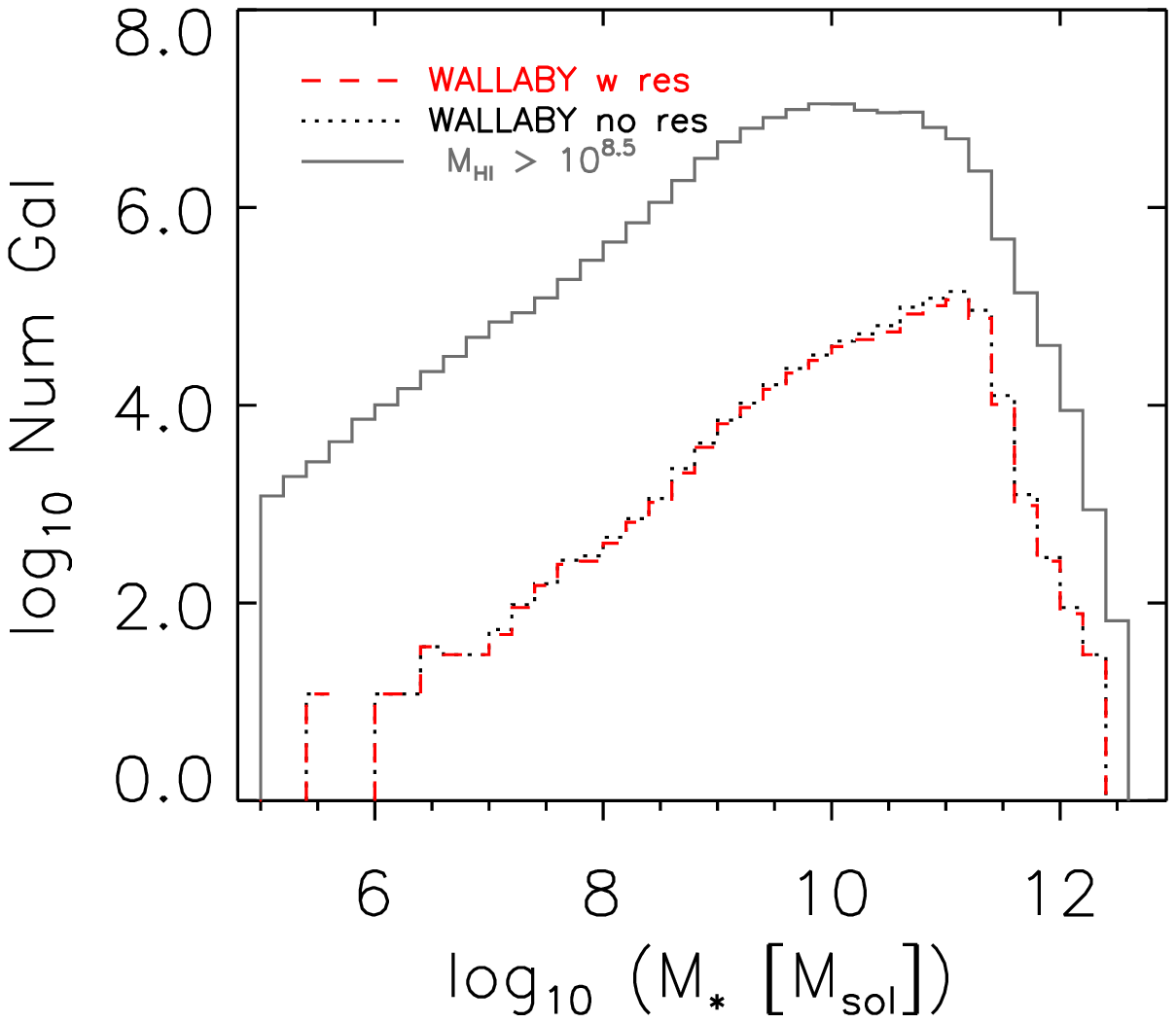, scale=0.45} \\
    \epsfig{figure=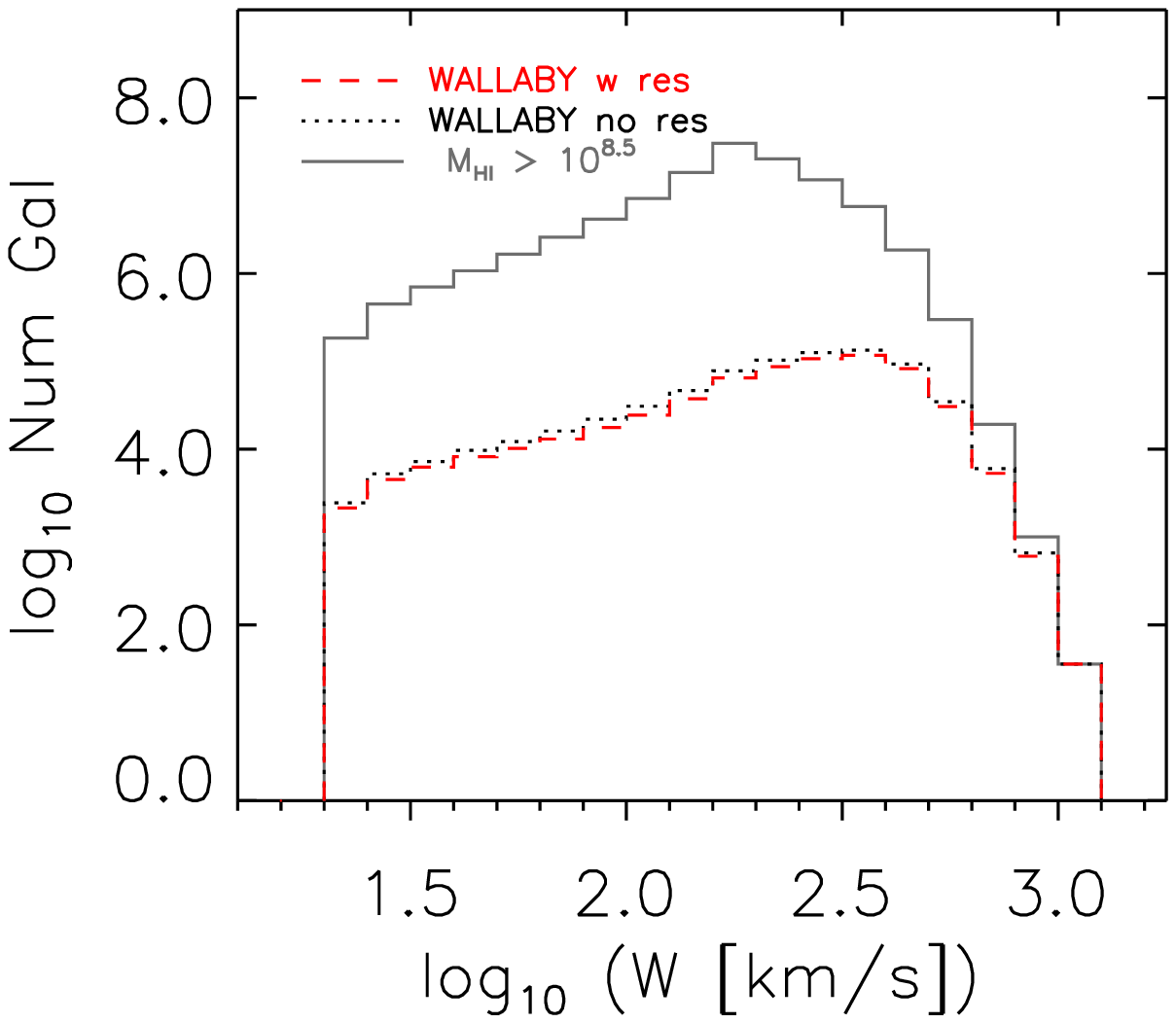, scale=0.45} \\
    \caption[Halo properties]
            {In these figures we consider the underlying galaxy distribution, from top to bottom, as a function of total halo mass,
	   HI mass, stellar mass and velocity width for a blind HI survey, WALLABY.
            As a result of the wide range in \HI - halo mass ratios probed, Fig.~\ref{fig:HI_MVir_ratio}, we expect the 
            ASKAP surveys to probe stellar and halo masses over 4 orders of magnitude, as well as a wide range of
            halo velocity widths. We repeat this figure for the case of DINGO DEEP in Fig.~\ref{fig:galprop_dingo}.}
    \label{fig:galprop_wallaby}
  \end{center}
\end{figure}

We now consider the type of galaxies that the WALLABY sample will contain in Fig.~\ref{fig:galprop_wallaby}, 
this represents the true strength of using complex simulations to predict the galaxy catalogue as 
one can quantify the underlying galaxy population that a blind \HI survey will probe.
We consider several key galaxy properties, from top to bottom panels these are the total 
halo mass, \HI mass, stellar mass and velocity width of the
galaxies. From this figure it is clear that WALLABY can expect to detect systems ranging over 4 orders of magnitude in halo virial mass
and 7 orders of magnitude in stellar mass. The velocity width is approximately proportional to the
square root of the mass enclosed and hence has a smaller dynamic range but we can expect to detect systems with velocities ranging from 
$20\,\rm km \,s^{-1}$ to $1000\,\rm km \,s^{-1}$.
This broad selection of galaxies will be a unique resource to astronomers interested in galaxy formation as well 
offering powerful constraints on the nature of dark matter through structural analysis using the velocity widths of 
these systems. 

\section{DINGO}
A great strength of ASKAP is the possibility of performing deep HI surveys to provide cosmologically
representative samples of HI rich galaxies beyond the local universe ($>100 k$ galaxies out to $z=0.43$, 
as given in Table~\ref{tab:survey_results}).
The deeper \HI surveys with ASKAP, termed DINGO, faces a unique challenge relative to the low redshift 
WALLABY survey, namely the greater risk of confusion of \HI sources at the increased distances which DINGO 
will probe relative to the case of WALLABY. We were able to ignore the issue of confusion for WALLABY (and 
have indeed checked that this assumption is valid) but must consider it for the deeper surveys before estimating 
the galaxy catalogues that DINGO will likely produce.

\subsection{Source Confusion}\label{sec:confusion}
A potential issue in any galaxy survey is the ability to uniquely identify detections rather than artificially blending overlapping 
sources, in the sky plane and along the line of sight, as one object. In previous work~\citep{Duffy:12b} we argued with simple 
analytic calculations that the incidence of confusion was slight when one had a spectroscopic survey. 
With our galaxy catalogue we can make a much improved estimate by utilising the actual on-sky angular diameter and velocity 
widths of the galaxies as well the realistic clustering of sources in space to determine whether they overlap as well as 
considering the improvements afforded by imaging with the 6km baselines of ASKAP. 

\begin{figure*}
  \begin{center}
    \begin{tabular}{cc}
   \epsfig{figure=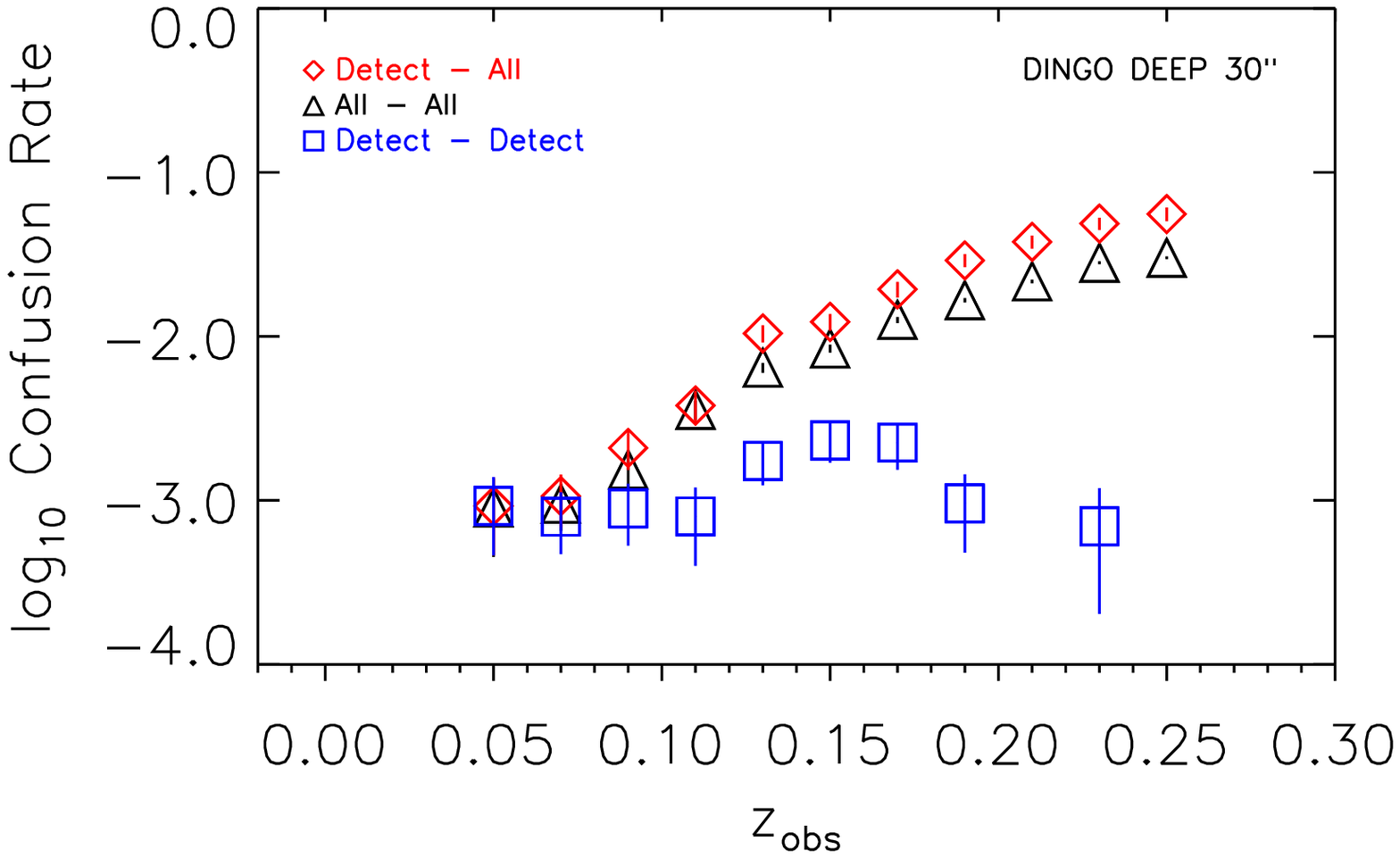, scale=0.45} &
   \epsfig{figure=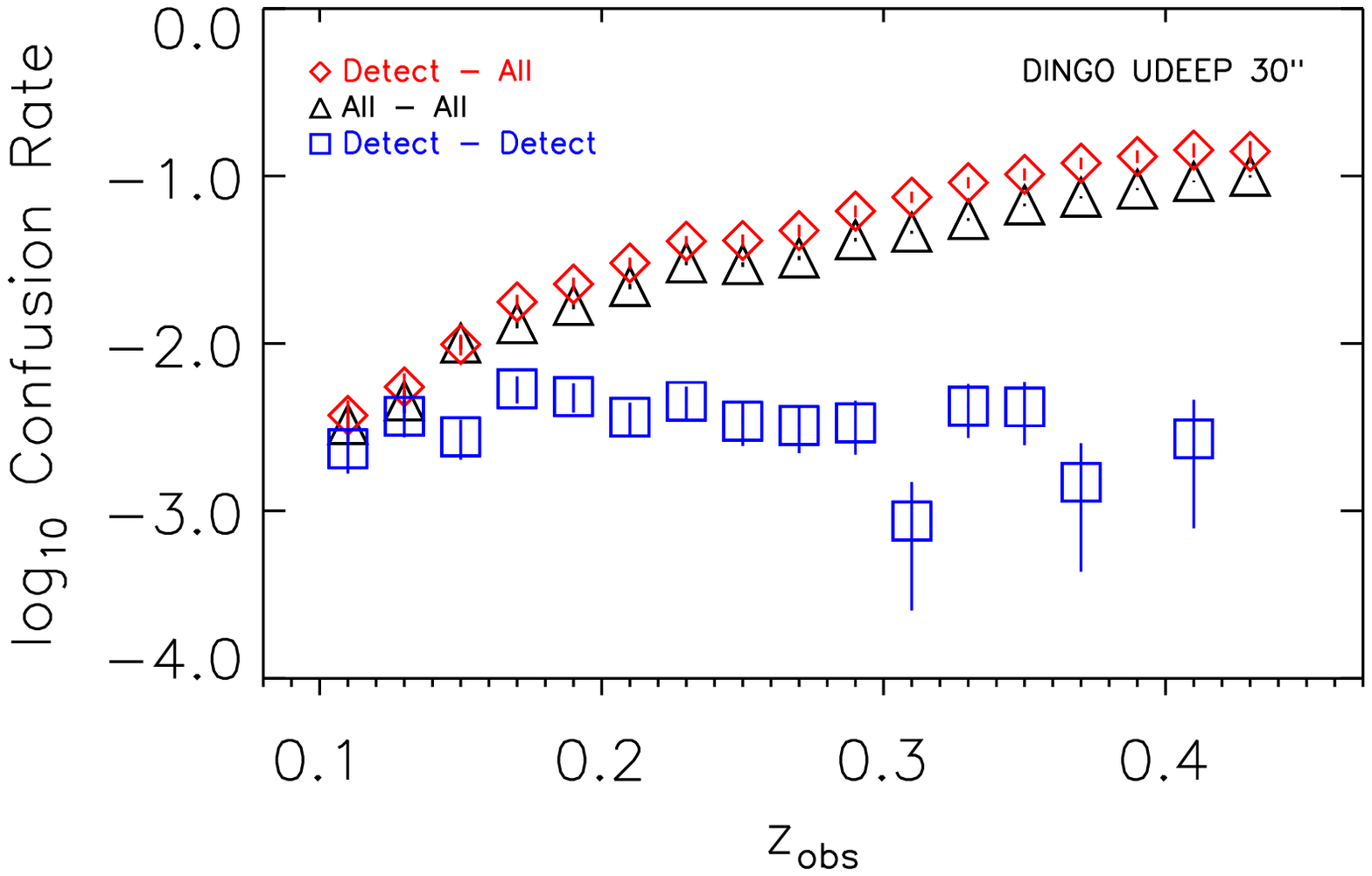, scale=0.45} \\
   \epsfig{figure=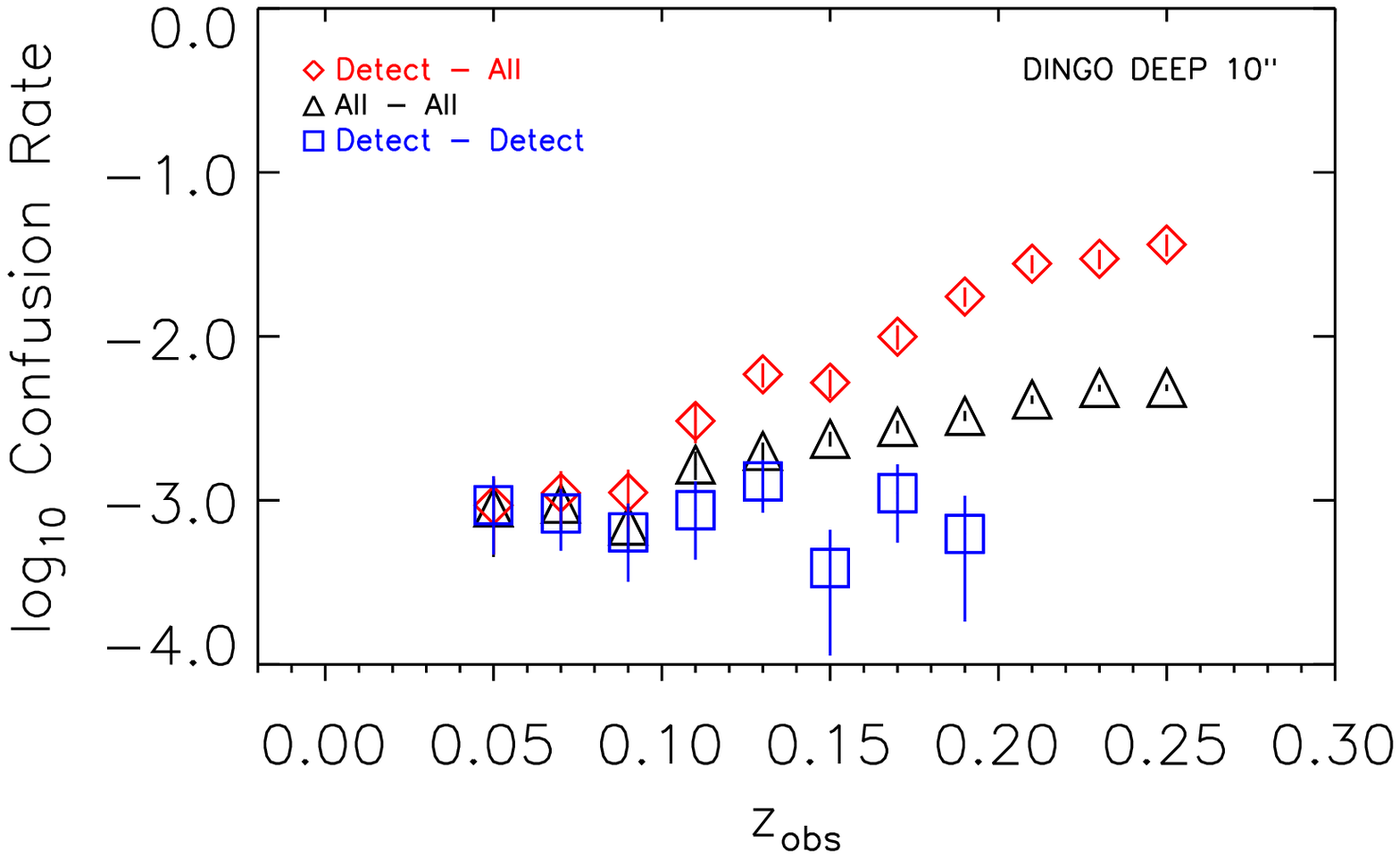, scale=0.45}  &
   \epsfig{figure=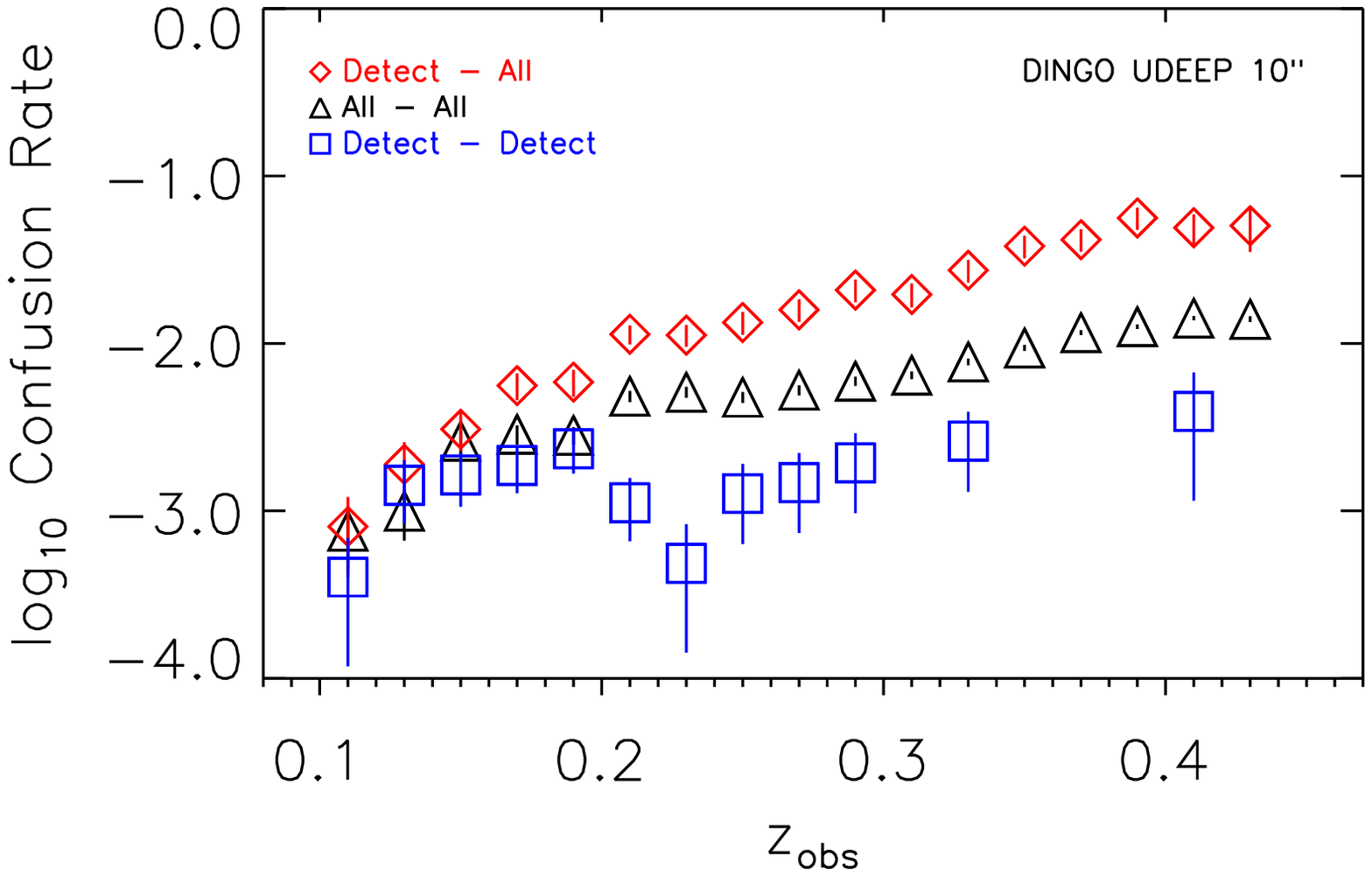, scale=0.45}  \\
\end{tabular}
    \caption[confusion rates for all galaxies]
            {We investigate the confusion rates of overlapping galaxies in both the DINGO DEEP and UDEEP surveys (left and right columns respectively) 
            with two possible dish configurations for ASKAP,  a core design with only 2km baselines and an extended 6km case (giving 30'' and 10'' resolution shown in the top and bottom rows respectively).
            To calculate the incidence rate of galaxies overlapping in redshift slices of width $\Delta$$z=0.02$ we utilised the angular extent of the galaxies on the sky 
            (if unresolved the angular diameter was fixed to be the telescope resolution) together with the velocity widths along the line of sight.
            We consider two cases of confusion; between all galaxies above a limiting \HI mass $M^{\rm lim}_{\rm \HI} = 10^{8.5} \Msol$, `All-All'
            in black triangles, and the confusion rate of the detected galaxies in the survey with any galaxies, `Detect-All' red diamonds. The former case is normalised
            by the number of galaxies in a redshift slice, the latter by the number of detections in the same slice. Errors are Poissonian. 
            The largest confusion rate is seen to be at the
            survey edge of the deepest survey, DINGO UDEEP (top right panel), which reaches $\sim$10\% if only 30'' resolution is attainable. If a strategy of 
            follow-up observations at 10'' are adopted then this rate becomes negligible ($<3\%$).
            Intriguingly, we note that the detected sources are more often confused than the average rate, this is discussed in the text.}
    \label{fig:confusion}
  \end{center}
\end{figure*}

In Fig.~\ref{fig:confusion} we consider the confusion rate, defined as the number of galaxies that overlap in angular projection (minimum on sky angle is set by the telescope resolution,
30'' or 10'') and along the line of sight divided by the total number of galaxies within a redshift slice $\Delta z = 0.02$. We note that we have an effective mass limit for this comparison of 
$M^{\rm lim}_{\rm \HI} = 10^{8.5} \Msol$ at all times. 
If we considered galaxies below this \HI mass limit, the confusion rate would naturally increase, yet little additional 
mass would be spuriously attributed to the main object\footnote{Provided the \HI mass function has a faint
end slope sufficiently shallow that the total mass in systems below a given threshold mass converges i.e. $\alpha > -2$, 
based on~\citet{hipass} who find $\alpha=-1.37\pm{0.03}\pm{0.05}$ measured down to $M_{\rm \HI} \approx 10^{7} \Msol$ which is sufficiently shallow for our conclusion to stand.} 
to a typical detection in DINGO DEEP or UDEEP which is over an order of magnitude more massive than this lower limit. 
The three curves in this plot are, respectively, the confusion rate between all galaxies above the lower \HI 
mass limit, relative to the total number of galaxies in the survey volume (termed `All-All'), the `Detect-All' case 
which considers the number of galaxies that are detected by DINGO DEEP or UDEEP that are also then 
overlapping with any galaxy above $M^{\rm lim}_{\rm \HI}$ normalised by the number of detected galaxies and 
finally `Detect-Detect' which is the self-confusion rate between detected galaxies. 

The DINGO DEEP survey is considered in the left column of Fig.~\ref{fig:confusion} for a resolution of 
30'' (10'') in the top (bottom) panel. This is repeated for the DINGO UDEEP survey is in the right-hand column.
Immediately apparent is that the confusion rate increases as a function of redshift as the number density of sources within a typical beam increases. The second point is that the confusion rate when one considers a detected subsample, `Detect-All', 
is typically $\sim 0.3$ dex higher than the `All-All' confusion rate of all galaxies above $M^{\rm lim}_{\rm \HI}$. 

This is at first a surprising result as often HI sources are often seen as the most weakly clustered 
population~\citep[e.g.][]{Meyer:07,Martin:12} and hence the `Detect' sample, which will be typically be more massive than `All' ,
should be just as confused as `All'. The $0.3$ dex increase is in fact a combination of three possibly effects.

The first is that more massive objects do lie within denser environments, and while \HI is a weakly correlated tracer there are 
tentative results that overdensities are still detectable over `field' galaxies, as shown for the case of the Fornax cluster in~\citet{Waugh:02}.
The second factor is that the semi-analytic model of~\citet{Croton:06} did not model proximity effects for infalling objects within 
overdense regions through mechanisms such as gas stripping, harassment or starvation. 
Therefore we should expect a higher clustering signal than otherwise
expected making our conclusions about confusion limits for the ASKAP surveys a conservative case.

The final factor is numerical in nature and is because we can only track objects to $M^{\rm lim}_{\rm \HI} = 10^{8.5} \Msol$. This means 
that although massive galaxies in our catalogue have resolved satellites, objects at this limit won't have their own satellites. 
We can test the severity of this issue by choosing a mass cut of $M^{\rm lim}_{\rm \HI} = 10^{9} \Msol$ and 
re-running the analysis, in this case the difference is lessened (as might be expected since proportionally more
galaxies above this limit in the volume are detected) but only by $0.1$ dex leaving the additional discrepancy a likely combination of
the first two explanations.

Within the entire DINGO DEEP sample of detections 766 (350) galaxies, or a rate of 1.4\% (0.7\%), are confused with galaxies of 
mass greater than $M_{\rm \HI} = 10^{8.5} \Msol$ with the ASKAP 30'' (10'') beam configuration. 
Of the detections that are confused with {\it other} detections, i.e. self-confused, the rate is of course far lower with only 62 (34) 
galaxies, or a rate of 0.11\% (0.07\%), with the ASKAP 30'' (10'') beam configuration. 
With a deeper galaxy sample one might expect the DINGO UDEEP survey to suffer a greater incidence of confusion within a fixed telescope beam 
resolution, and indeed 3971 (1015) galaxies, 
a rate of 4.9\% (1.3\%), suffer confusion for the 30'' (10'') ASKAP beam. 
The self-confusion incidence in UDEEP is also slightly higher than in DEEP, with 406 (118) galaxies
overlapping with other detected galaxies, a rate of 0.5\% (0.15\%). 

\subsection{HI detections}
\begin{figure}
  \begin{center}
    \epsfig{figure=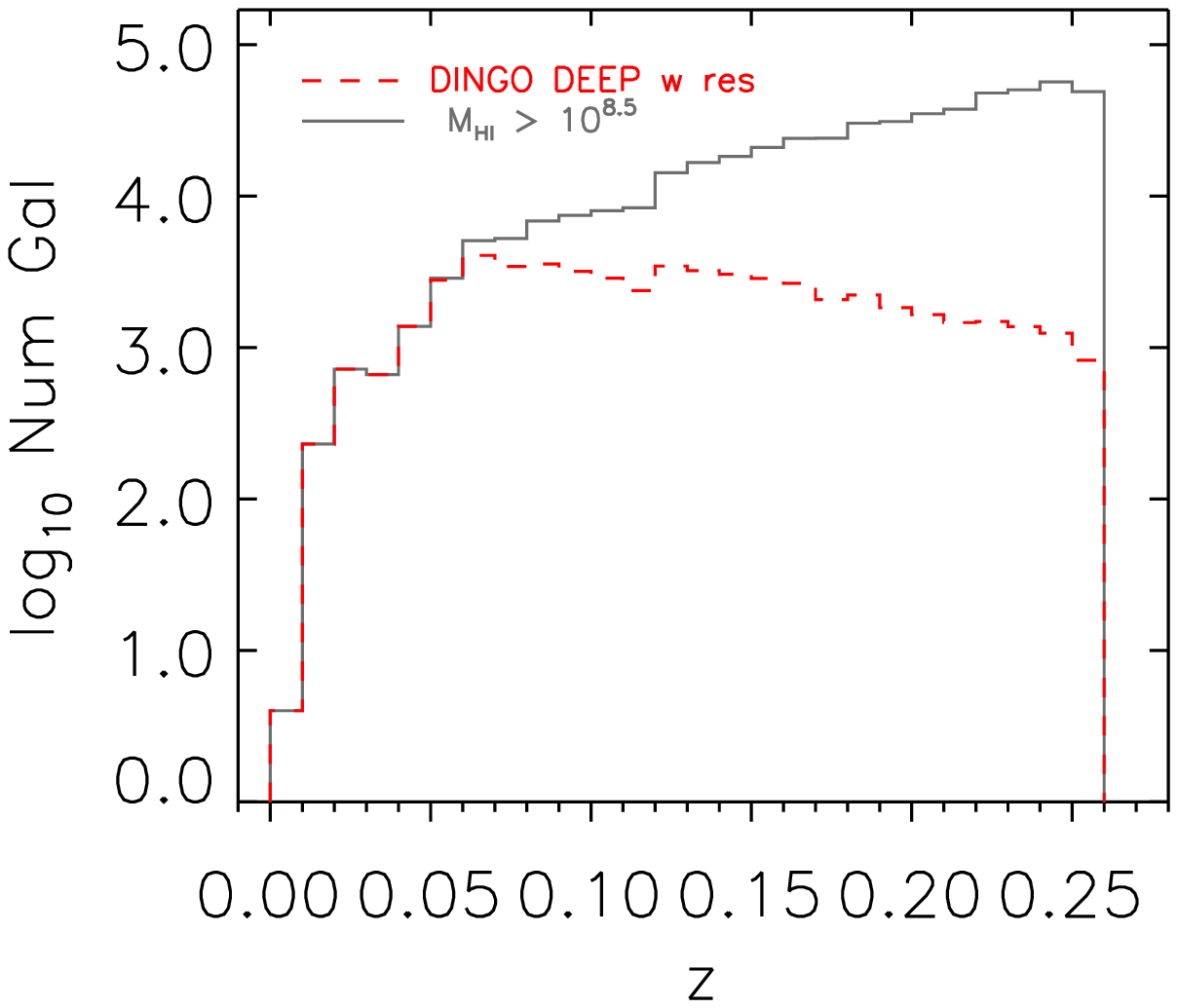, scale=0.5}
    \epsfig{figure=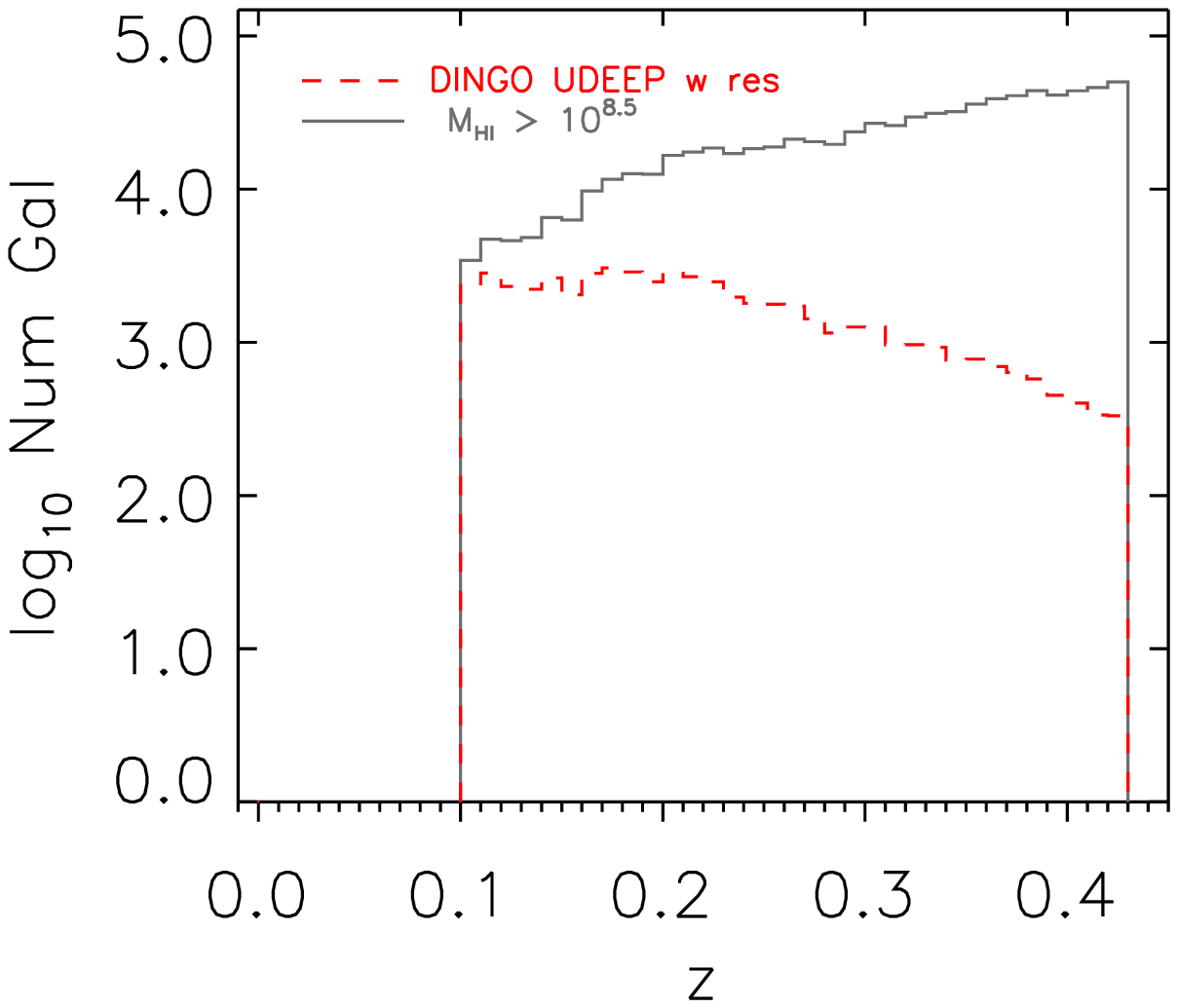, scale=0.5}  
    \caption[Survey]
            {The predicted galaxy number counts for the surveys; DINGO DEEP and UDEEP
            (top and bottom respectively). 
            The grey solid histograms represent the initial galaxy catalogue and the red dashed indicate source finders that only
            incoherently summate signal from extended sources and hence is a more conservative estimate of the ultimate telescope 
            performance.
            These figures demonstrate the strength of ASKAP 
            to probe the evolution of \HI out to $z = 0.26$ and $0.43$ in the DINGO DEEP and UDEEP surveys, respectively.}
            \label{fig:dndz_dingo}
  \end{center}
\end{figure}

With the previous result that the overall number counts in DINGO will be only slightly affected by the issue of source confusion we can 
consider the distribution of the detections as a function of redshift in Fig.~\ref{fig:dndz_dingo} for the DEEP and UDEEP aspects of 
DINGO.
We see immediately that the surveys compliment each other, with the increased observing time over smaller areas enabling the UDEEP 
part of DINGO to extend the \HI detections to $z=0.43$. Note that there are several orders of magnitude more galaxies undetected 
in the survey volume with $M_{\rm \HI} = 10^{8.5} \Msol$. These can be stacked to create a measurable signal from individually 
undetected sources (after cross-correlating with optical surveys, the issue of identifying these galaxies is considered in 
Section~\ref{sec:misidentification}).

\begin{figure}
  \begin{center}
    \epsfig{figure=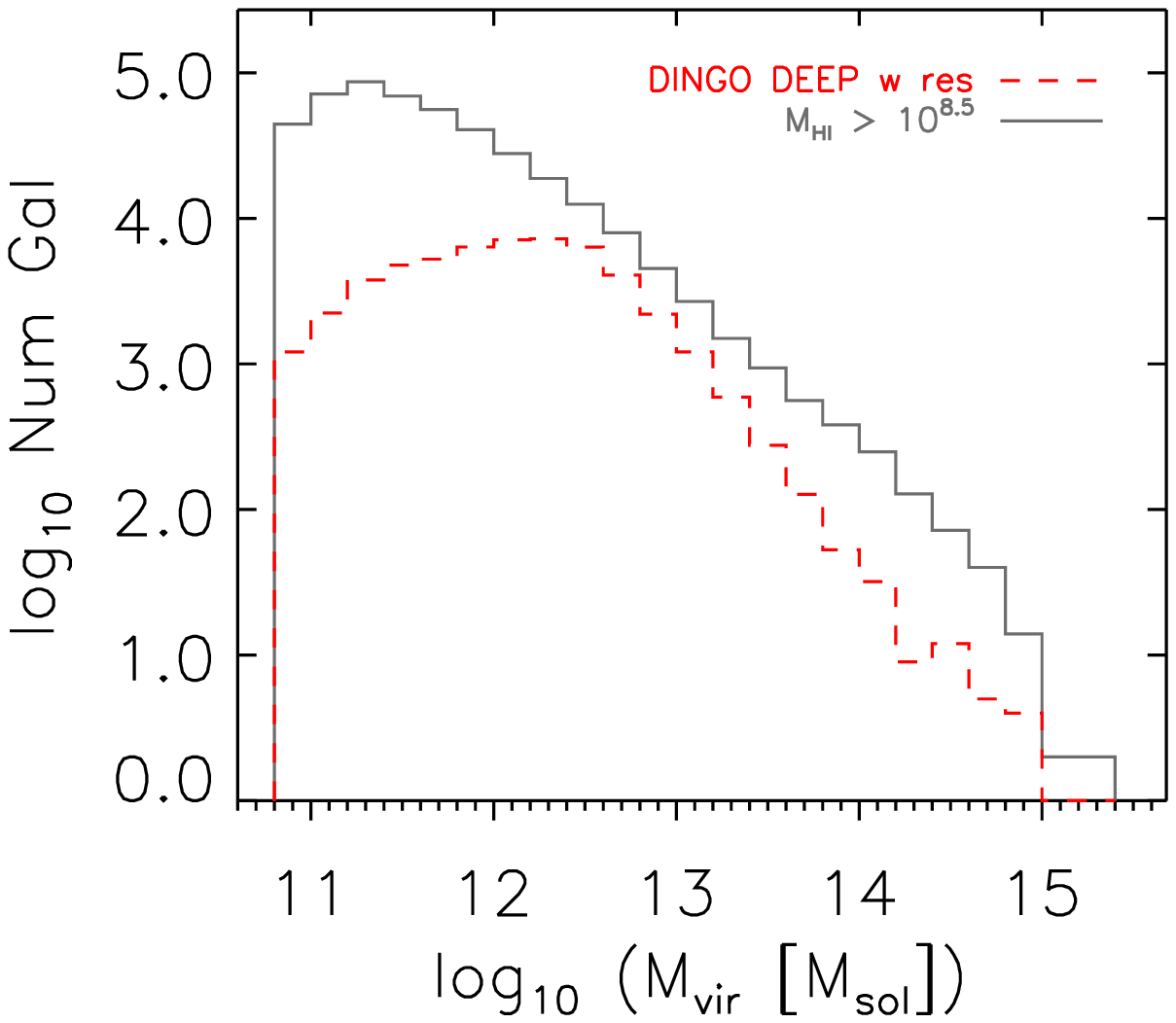, scale=0.45} \\
    \epsfig{figure=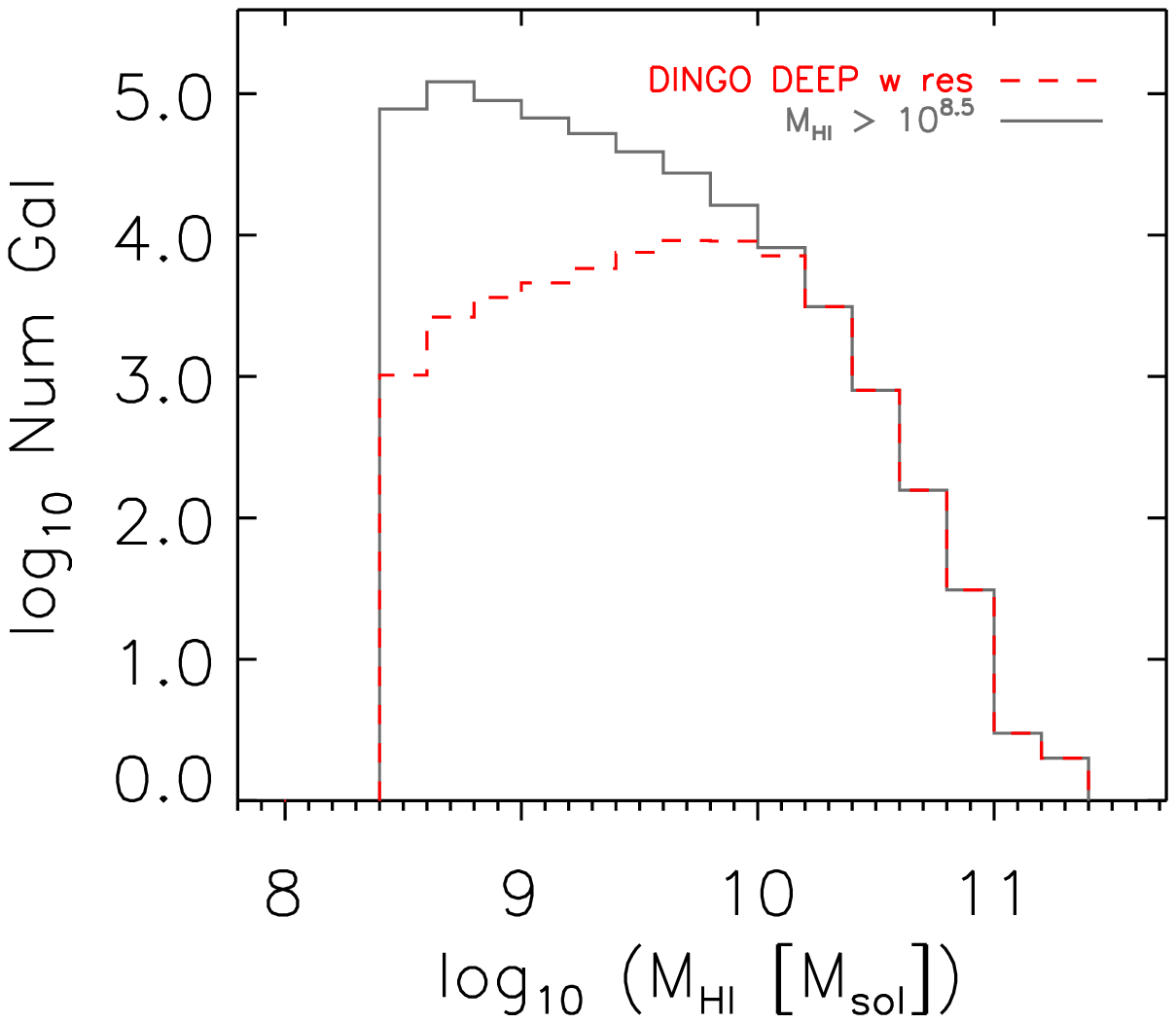, scale=0.45} \\
    \epsfig{figure=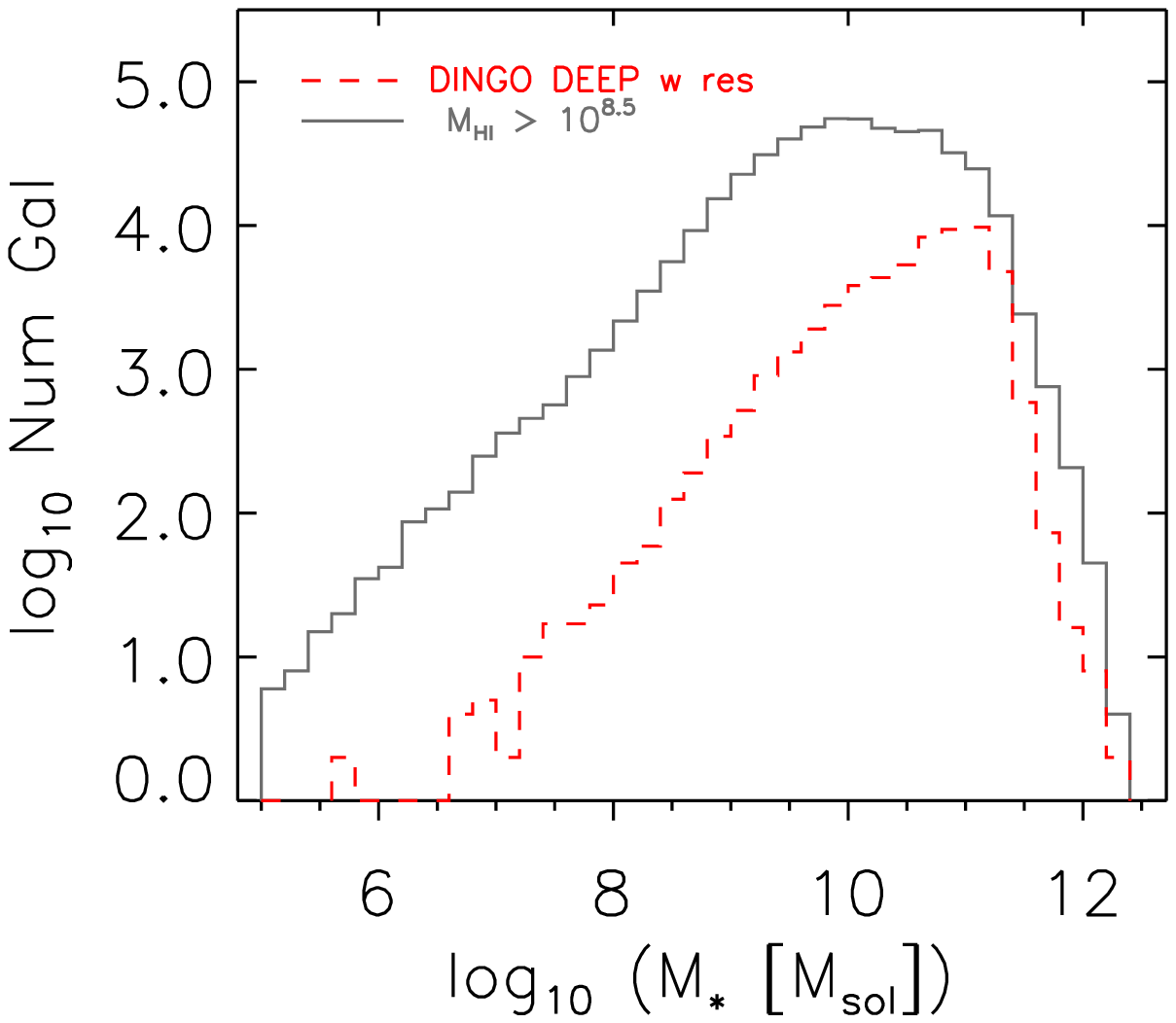, scale=0.45} \\
    \epsfig{figure=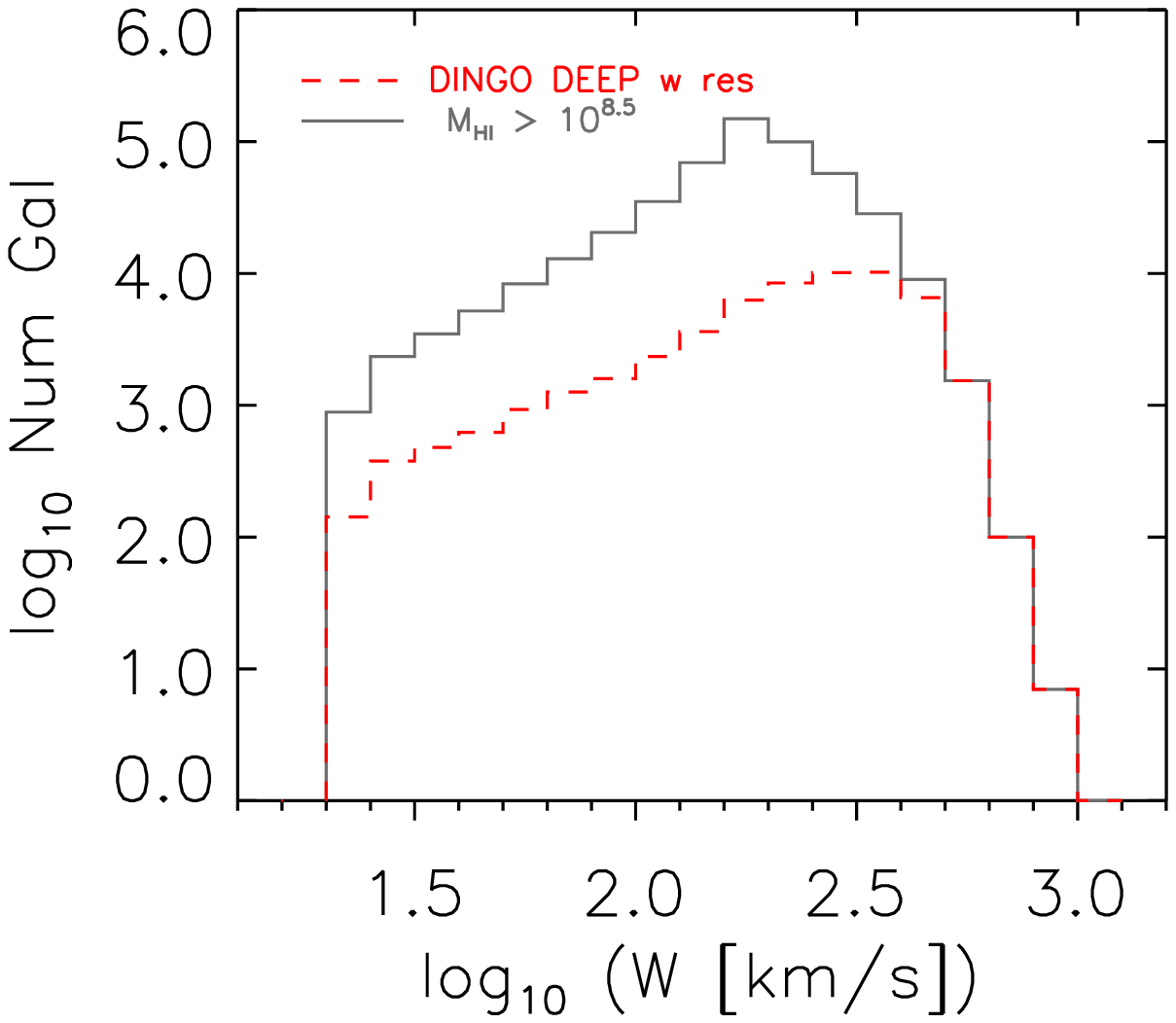, scale=0.45} \\
    \caption[Halo properties]
            {In these figures we consider the underlying galaxy distribution, from top to bottom, as a function of total halo mass,
	   HI mass, stellar mass and velocity width for a deep HI survey, DINGO DEEP.
            The range of properties probed is similar to WALLABY, given in Fig.~\ref{fig:galprop_wallaby} previously. }
    \label{fig:galprop_dingo}
  \end{center}
\end{figure}
The properties of the detected galaxies in DINGO are similar to that probed by WALLABY, with the caveat that the simulations
only contain galaxies with $M_{\rm \HI} \ge 10^{8.5} \, {\rm M}_{\odot}$ and hence the likely low mass systems in the latter are missing.
In Fig.~\ref{fig:galprop_dingo} we consider the total halo mass, HI mass, stellar mass and intrinsic rotational width of the sample of
galaxies detected by DINGO DEEP. The survey detects 4 orders of magnitude in total mass, 7 orders of magnitude in stellar mass
and nearly 2 orders of magnitude in velocity width.

\begin{figure}
  \begin{center}
   \epsfig{figure=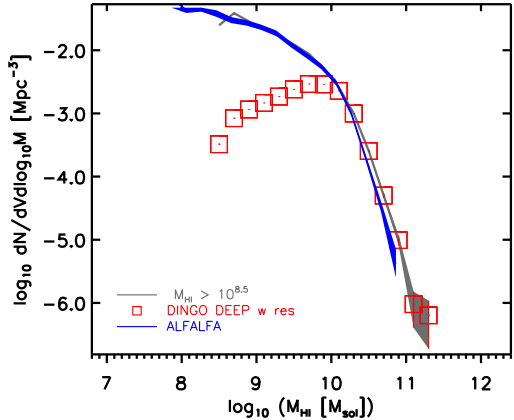, scale=0.5} 
    \caption[HI mass function]
            {The blue line is the observed \HI mass function from ALFALFA~\citep{Martin:10} with errors represented by the thickness of the 
            line and in grey is the \HI mass function from the simulation volume, with Poissonian errors given by the shaded region. 
            We then perform an observation using the DINGO DEEP survey parameters
            and plot the resultant {\it flux limited} mass distribution recovered in the red squares. Error bars are Poissonian. 
            DINGO DEEP is complete for galaxies above the `knee', $M_{\rm \HI} \ge 10^{10} \, {\rm M}_{\odot}$, allowing us to measure
            evolution in the high mass end of the \HI mass function over the entire redshift range of the survey; and to progressively
            lower redshift ranges for lower masses.}
    \label{fig:himassfn}
  \end{center}
\end{figure}

A significant science case for large area \HI surveys is to measure the \HI mass function; as is shown in Fig.~\ref{fig:himassfn}, 
the ASKAP telescope will create a complete mass function above $M_{\rm \HI} \approx 10^{10}\, {\rm M}_{\odot}$ with the DINGO survey.
This will enable the measurement of evolution in the high mass end of the mass function across the entire redshift range, and 
to progressively lower masses for lower redshift ranges.

\begin{figure}
  \begin{center}
    \epsfig{figure=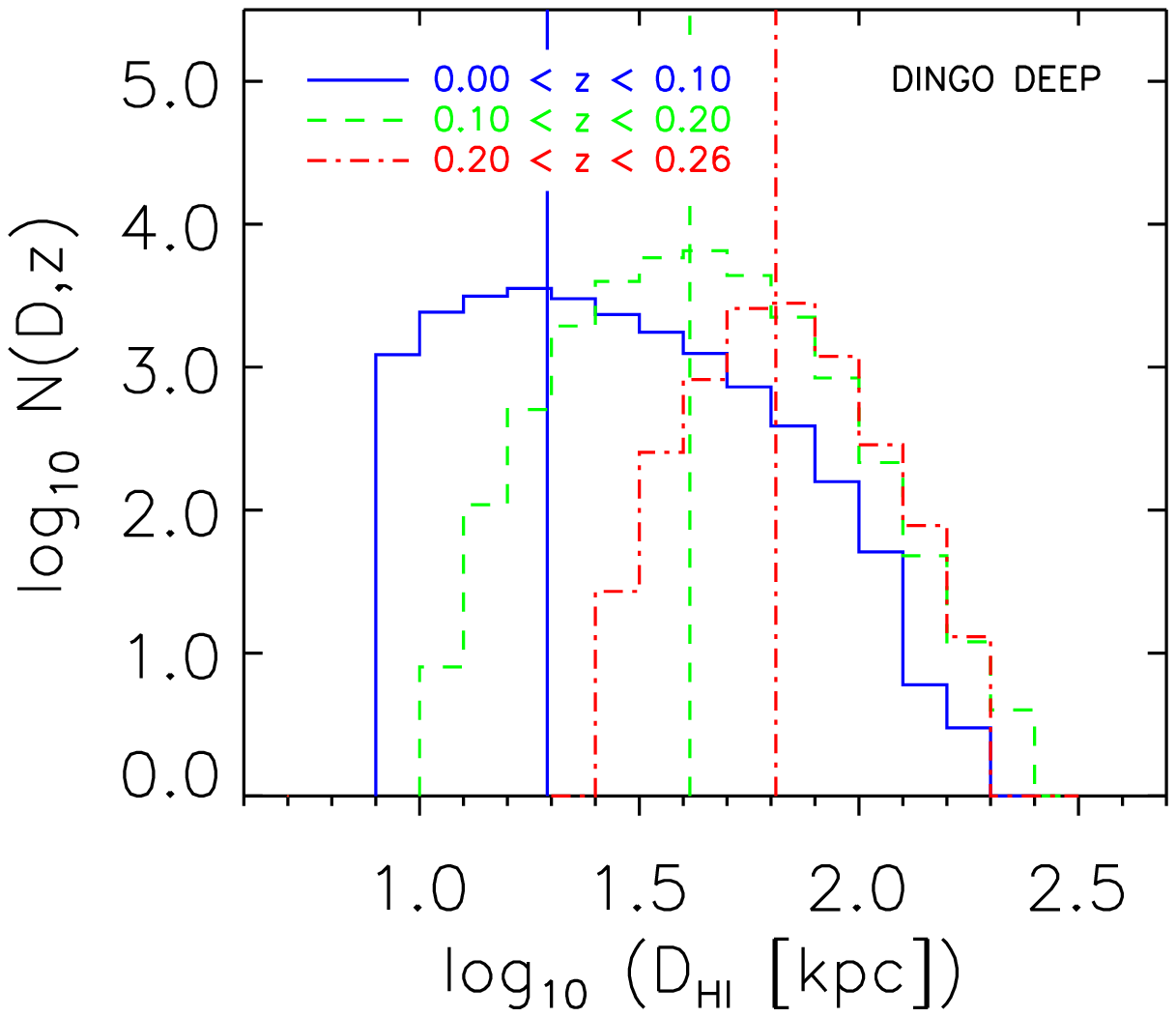, scale=0.5} 
    \epsfig{figure=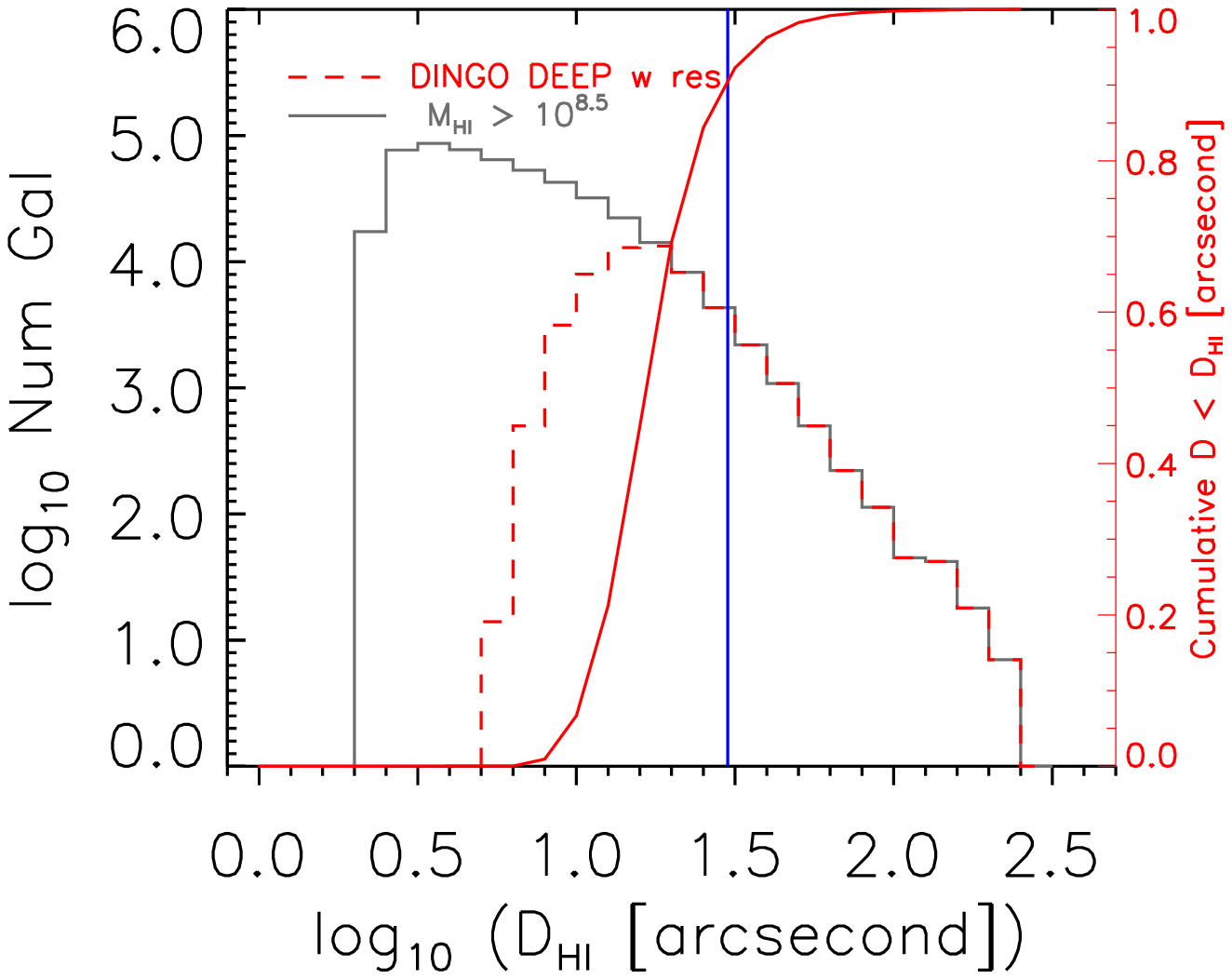, scale=0.5} 
    \epsfig{figure=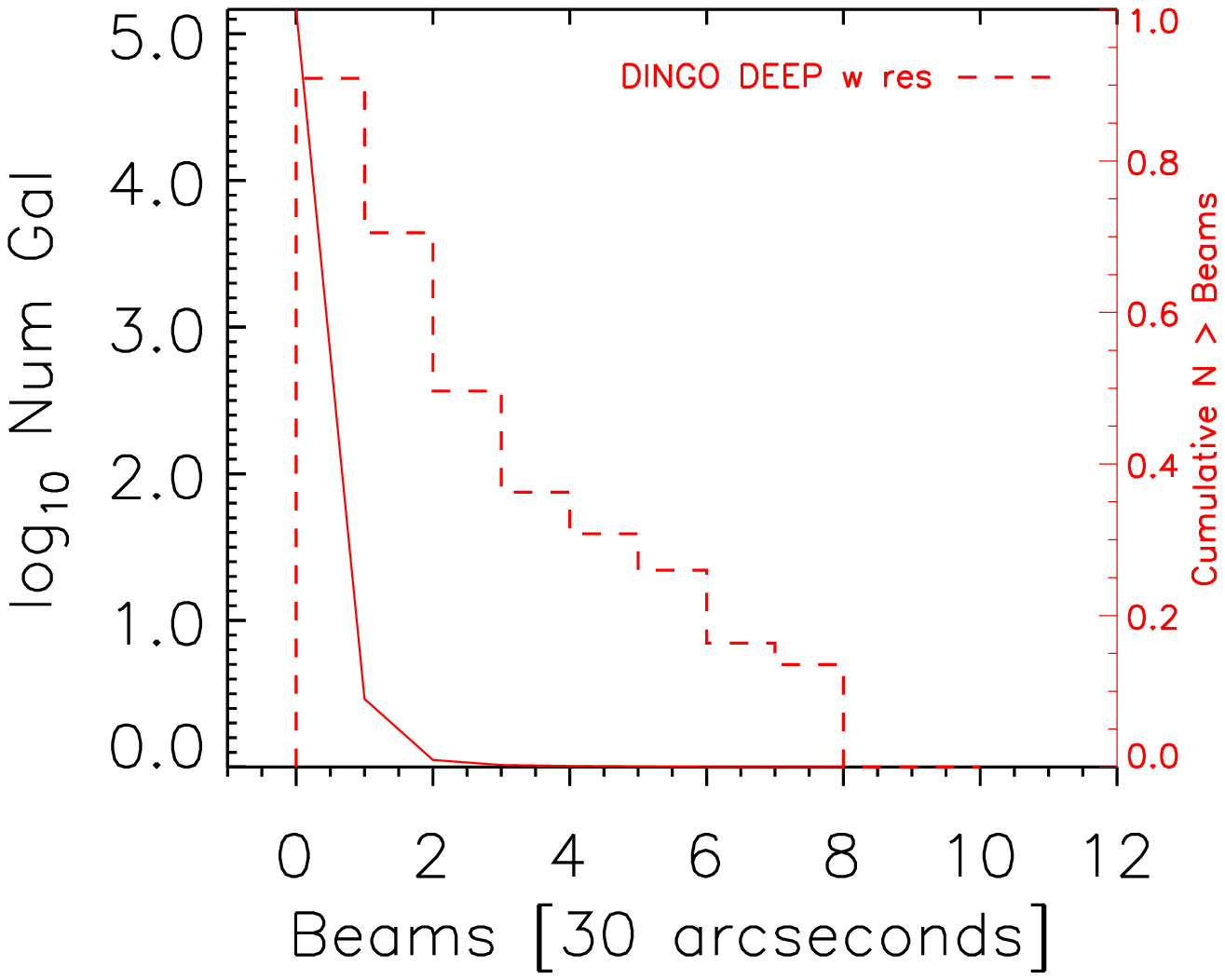, scale=0.5} 
    \caption[Galaxy Resolved]
            {In these figures we consider the sizes of the galaxies as a function of intrinsic physical diameter as estimated using 
            Eqn.~\ref{eq:size mass relation} for several redshift cuts (top), angular extent on the sky (middle) and the number of beams that 
            the galaxies will be resolved by (bottom panel) for the proposed DINGO DEEP survey.
            For the case of DINGO UDEEP the majority of galaxies, even with the resolution afforded by the ASKAP core
            configuration will be only marginally resolved in angular extent at best hence we do not consider it here.}
    \label{fig:dhi_dingo}
  \end{center}
\end{figure}

We now consider the physical extent of the sources detected by DINGO in Fig.~\ref{fig:dhi_dingo} which are typically of smaller physical
extent (top panel) at a given redshift than the corresponding WALLABY sources as a result of the greater integration time per field which 
allows fainter (and hence physically smaller) galaxies to be detected. As shown in the middle panel of Fig.~\ref{fig:dhi_dingo} the 
majority of galaxies are unresolved (comparing the intersection of the cumulative red curve with the ASKAP $30\arcsec$ beam denoted 
by the vertical blue line).
If we create a histogram of the number of synthesised beams the galaxies are resolved by (bottom panel) we see less than a few percent of 
galaxies are {\em just} resolved by more than one beam (DEEP and UDEEP resolve $\le$9\% and 1\% of galaxies, respectively). 
We consider the case of observing these systems with the $10\arcsec$ synthesised beam when using 6km baselines, as shown in 
Table~\ref{tab:survey_values}, and find that now 95\% of objects are marginally resolved. Although not a key science case for DINGO
there could be as many as 100 galaxies resolved by more than 10 beams which will be a valuable dataset to compliment the 
WALLABY sample.

We illustrate the strength of DINGO as a deep survey instrument in Fig.~\ref{fig:lightcone} in which the large scale structure is 
clearly visible out to the edge of the survey at $z\sim 0.43$ allowing the \HI properties of galaxies to be studied as a function of both 
environment and redshift. 
These lightcones are presented as one contiguous field while in reality the DINGO fields are distributed across the sky,
thereby reducing the effects of cosmic variance. 

\subsection{Matching Optical Catalogues}\label{sec:misidentification}
We can use our catalogue to estimate typical misidentification rates when uniquely assigning optical counterparts to our \HI detections. This measurement differs slightly from the confusion rate discussed before. For the case where there are, say, two (three) possible 
counterparts one can always blindly choose one for the optical counterpart, statistically speaking, in this case we will be correct 
50\% (33\%) of the time. If one considered priors such as the size of these systems then we can be substantially more reliable than simple
blind guessing. Therefore the misidentification rate is slightly lower than the confusion rate due to the possibility of randomly assigning objects from the list of possible counterparts.

In~\citet{Duffy:12b} we considered two
optical catalogues, one with photometric redshifts (a typical redshift error for such galaxies is $\Delta z = 0.05$ as argued by~\citealt{Hildebrandt:08}) and a spectroscopic redshift
sample. We estimated that in the latter case, the uncertainty as to which galaxy was the counterpart was when two galaxy rotation widths overlapped, 
hence we {\it conservatively} assumed $\Delta z=0.002$ (twice the width of an $M^{\star}$ galaxy which is the likely system detected by ASKAP at high redshift\footnote{Although individual velocity widths of the galaxies are available, and indeed were used in the confusion study of 
Section~\ref{sec:confusion}, we adopt the limiting spectroscopic width in accordance with previous estimates from~\citet{Staveley-Smith:08}
and~\citet{Duffy:12b}.}). 

\subsubsection{DINGO DEEP}
In Fig.~\ref{fig:misidentification} we consider the case for the DINGO DEEP (UDEEP) survey in the left (right) columns. For both 
telescopes we consider a resolution of 30'' in the top panel and the high resolution imaging at 10'' in the bottom panel. For each case
we consider an optical catalogue with only photometric redshifts ($\Delta z = 0.05$) and one with spectroscopic redshifts 
($\Delta z = 0.002$), results given by the red square and black diamond points respectively. 
We find that the \HI galaxies surveyed with the 30'' ASKAP beam at $z=0.05$ can be uniquely 
identified with a single optical counterpart, with photometric errors, more than 96\% of the time (top left panel). 
However, the misidentification rate strongly rises to $\sim 18\%$ at the survey edge, $z=0.26$. If one has access to a spectroscopic
survey then even at the edge of the survey optical counterparts can be correctly identified 95\% of the time.

If one has access to increased resolution at 10'' for DINGO DEEP (bottom left panel) then the misidentification rate is $\le 3\%$ for the entirety 
of the survey even when only photo-z catalogues are available, in agreement with our conclusion in~\citet{Duffy:12b}.

\subsubsection{DINGO UDEEP}
As we showed for the confusion rates of DINGO UDEEP (see Section~\ref{sec:confusion}) it is much harder to separate galaxies at higher 
redshift than at low redshift and it is therefore unsurprising that the misidentification rate with an optical catalogue is also much higher than in 
the case of the lower redshift DINGO DEEP. 

\begin{figure*}
  \begin{center}
    \begin{tabular}{cc}
       \epsfig{figure=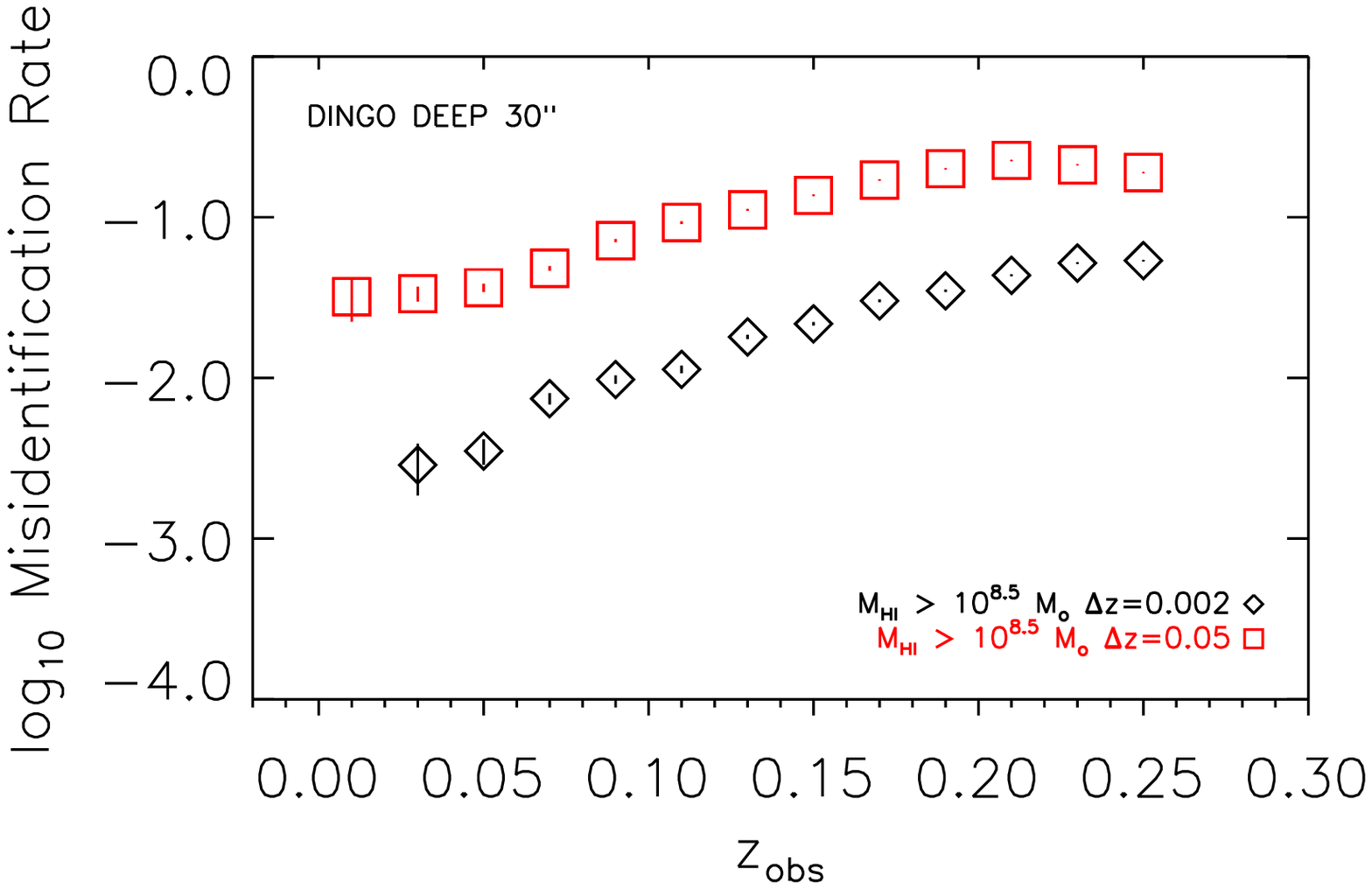, scale=0.45} &
       \epsfig{figure=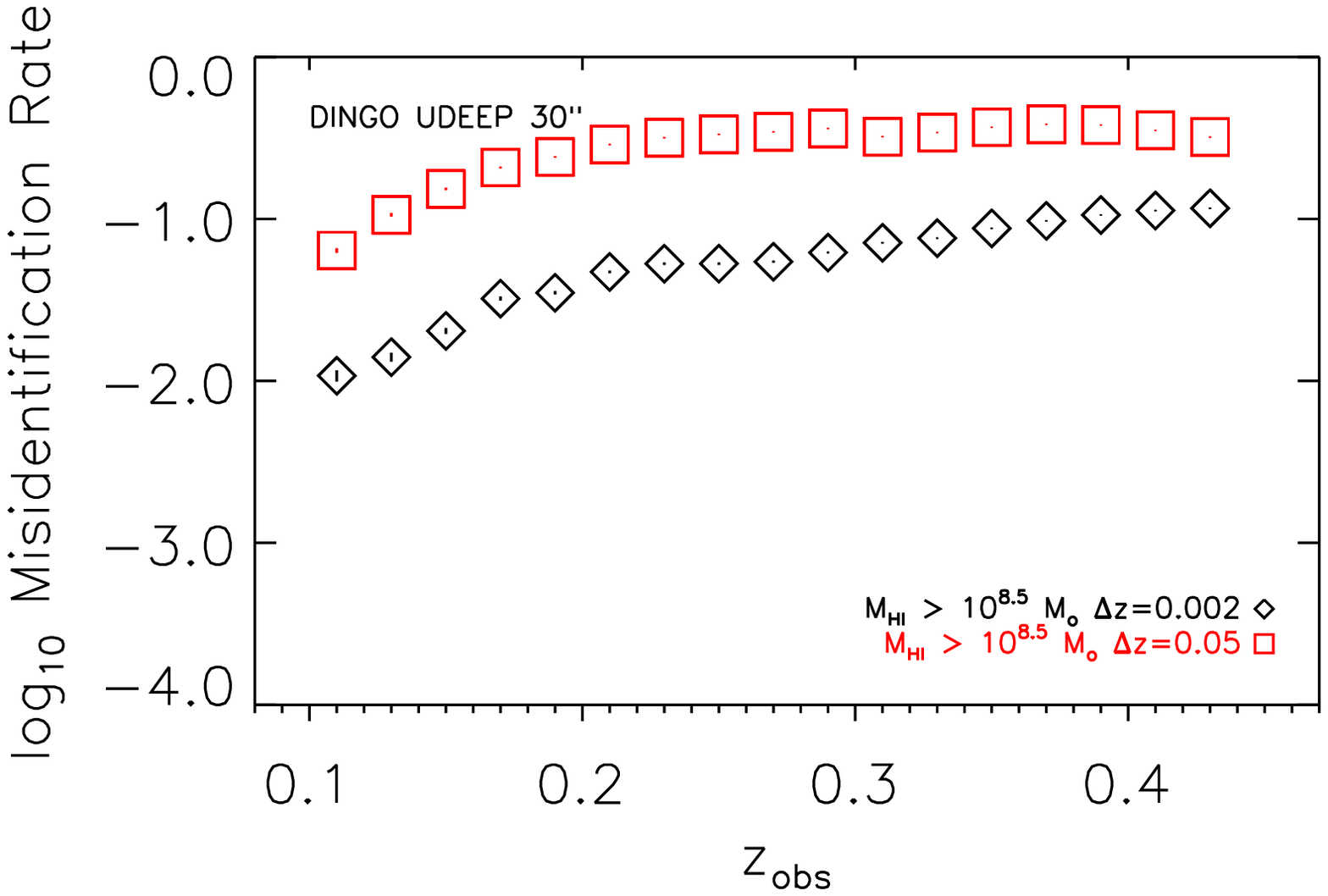, scale=0.45} \\
   \epsfig{figure=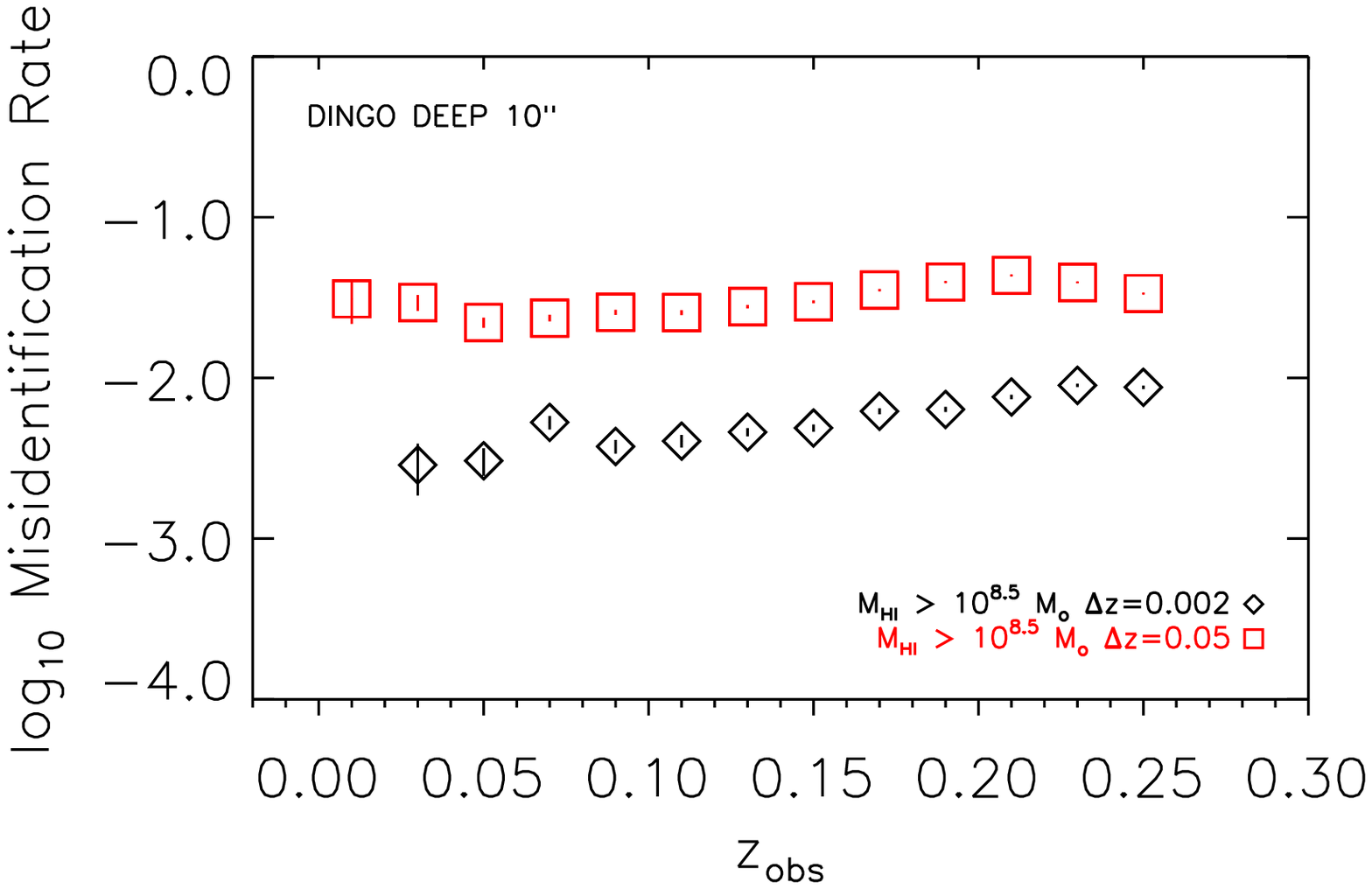, scale=0.45} &
   \epsfig{figure=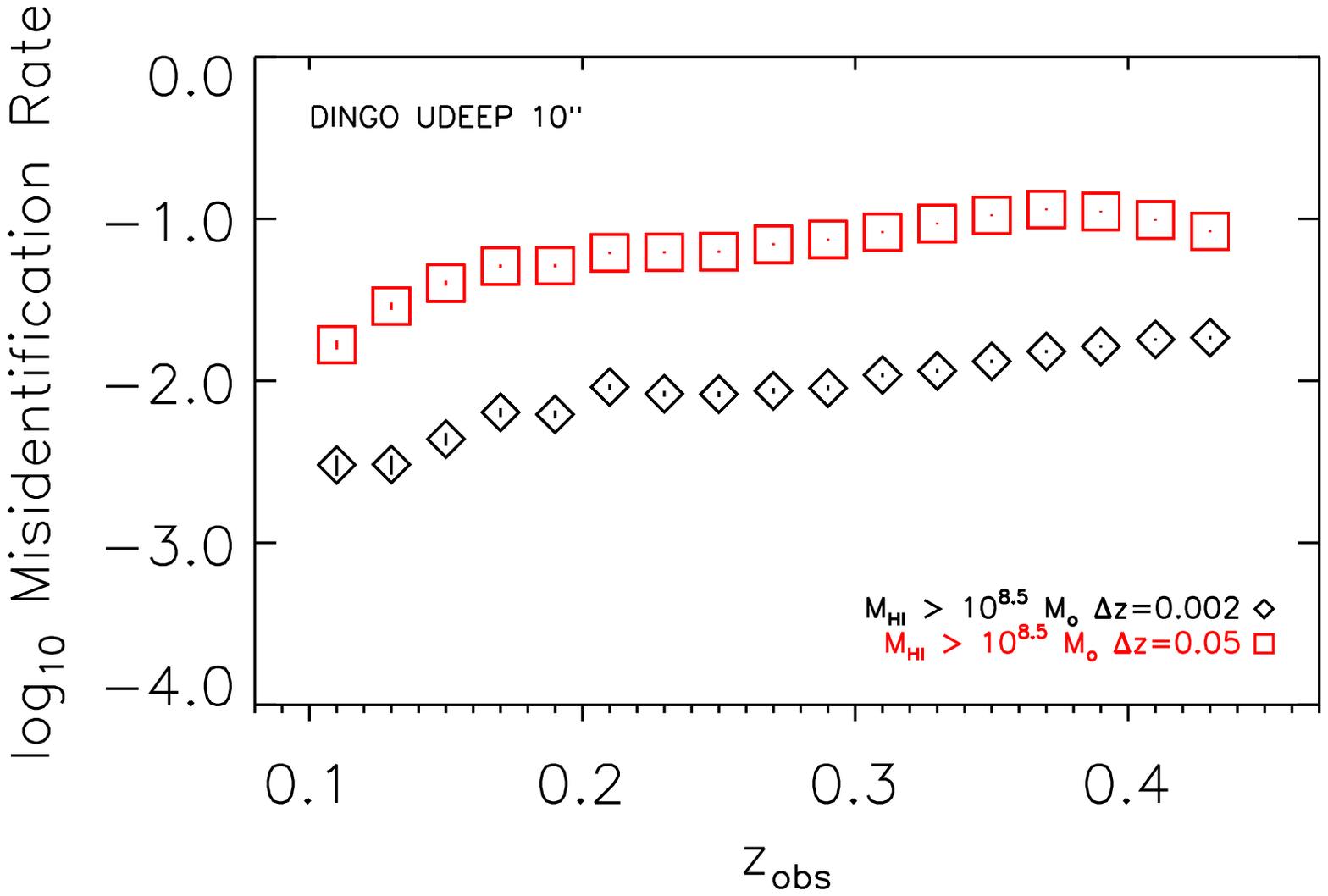, scale=0.45}  \\
\end{tabular}
    \caption[misidentification rates for all galaxies]
            {We have calculated the misidentification rate of galaxies when assigning an optical counterpart to \HI detections for the 30'' and 10'' resolutions of ASKAP (top and bottom 
            panels) in the DINGO DEEP and UDEEP surveys (left and right columns respectively). We considered two typical redshift errors that an optical catalogue would suffer; the first
            in red squares is a photometric redshift uncertainty of the position of 
            the optical counterpart position (as argued by~\citealt{Hildebrandt:08} this is typically $\Delta$$z=0.05$) and in black
            diamonds we consider a spectroscopic survey with typical uncertainties assumed to be twice the velocity width of an $M^{\star}$ galaxy ($\Delta$$z=0.002$) which 
            will be detectable throughout the redshift range of the survey~\citep{Duffy:12b}. The rates are normalised by the number of galaxies within the redshift slice, with Poissonian
            errors for each point. Within the greater volume of the ASKAP beam and optical photo-z errors there are more
            potential optical counterparts. At the edge of the DINGO UDEEP survey more than 30\% of galaxies will be misassigned. If there are follow-up 
            observations at 10'' the misidentification rate drops to $\sim 3\%$ ($\le 10\%$) for DINGO DEEP (UDEEP).}
    \label{fig:misidentification}
  \end{center}
\end{figure*}

We find that if one only has access to the 30'' DINGO UDEEP source list and attempt to uniquely identify counterparts from a photometric 
catalogue (red squares top right panel of Fig.~\ref{fig:misidentification}) then at the beginning (end) of the survey the misidentification rate
is 7\% (31.5\%). If using a spectroscopic optical catalogue (black diamonds top right panel of Fig.~\ref{fig:misidentification}) 
then the rates are 1-12\% across the survey.

As before if one images the field at 10'' (bottom right panel of  Fig.~\ref{fig:misidentification}) then even with a photometric catalogue the
misidentification rate is never higher than 6\% (reached at $z=0.43$) and when combined with a spectroscopic optical survey this would
drop to of order a percentage.

In conclusion, the rates of confusion for \HI galaxies are negligible
with ASKAP provided one has access to the highest possible resolution of 10'' synthesised beams. Key science 
goals that rely on the identification of the optical counterparts to \HI sources can make use of photometric catalogues, suffering from less 
than $9\%$ misidentification rates (even out to $z=0.43$ for the case of DINGO UDEEP) if one has the high resolution imaging.

If ASKAP synthesised beams of 30'' are used then DINGO DEEP can still identify an optical counterpart at $z= 0.05\, (0.26)$ using a photo-z 
catalogue more than $96\, (82.5)\%$ of the time. For DINGO UDEEP the photo-z catalogue will suffer a misidentification rate of $5\, (30)\%$
at the start (end) of the survey. If a spectroscopic catalogue is available then even with 30'' resolution the UDEEP sources will be uniquely
identified at $z=0.43$ 88\% of the time.

In conclusion, we find that the quality of the optical catalogue depends strongly on the ASKAP resolution. For DINGO DEEP and UDEEP 
science goals that can accommodate misidentification rates of order 5\% then a photometric redshift catalogue is sufficient with 10'' resolution  
imaging from ASKAP. 
However, for certain science goals such as \HI spectral stacking, spectroscopic optical catalogues are essential, irrespective of ASKAP 
resolution.

\section{Comparison of Surveys}\label{sec:compare}
\begin{figure}
  \begin{center}
   \epsfig{figure=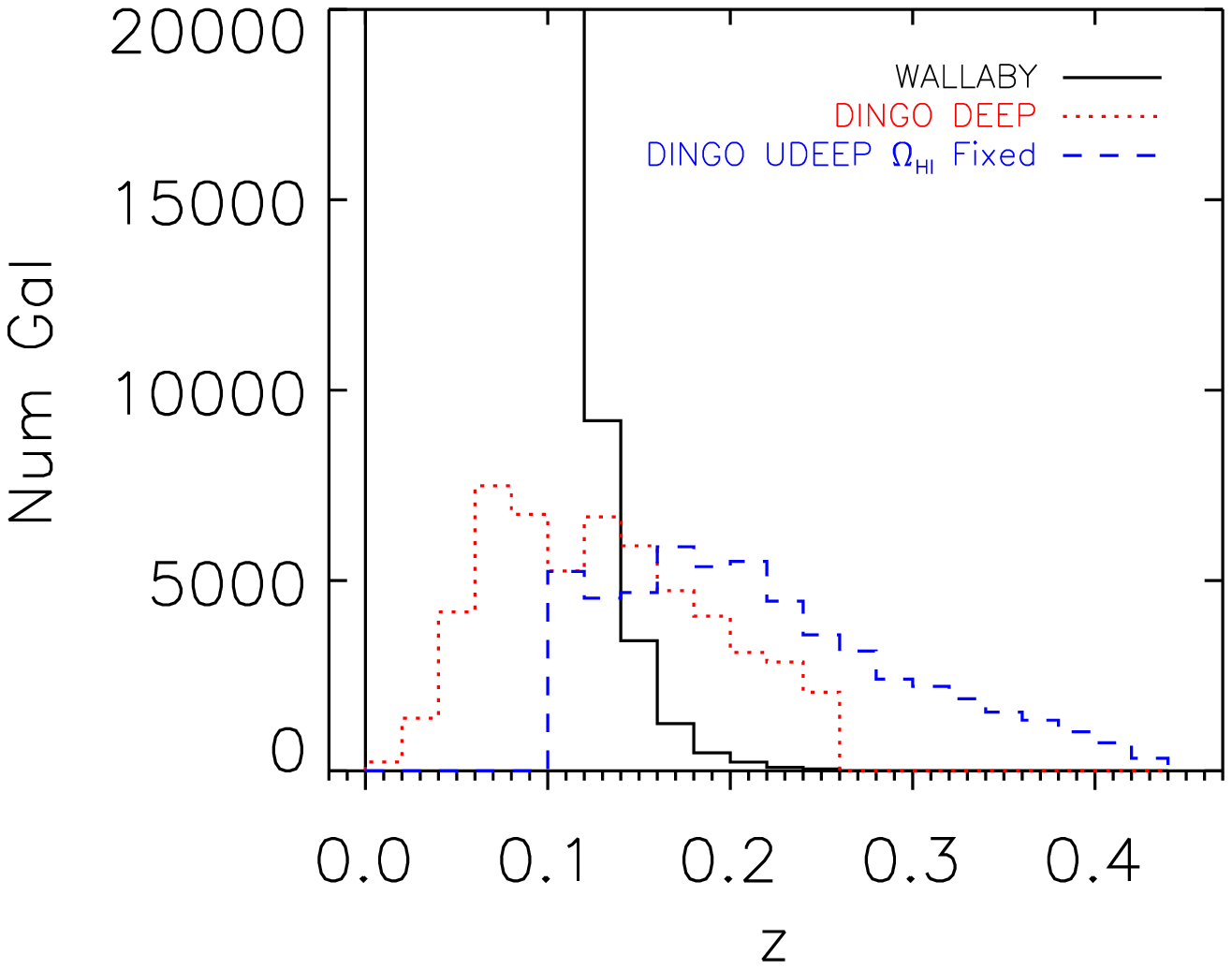, scale=0.5} 
    \caption[dndz]
            {The expected galaxy number density in redshift bins of width $\Delta = 0.02$ for the DINGO DEEP, UDEEP and WALLABY surveys. For DINGO UDEEP
            we take the conservative case that there is no evolution in the \HI mass function across the redshift probes, and for WALLABY we reduce
            the number of galaxies found by an order of magnitude for ease of comparison with the deeper surveys. Note the `wedding cake' design of these
            overlapping surveys which ensure significant numbers of galaxies are probed at each redshift.}
    \label{fig:dndz_all}
  \end{center}
\end{figure}

We combine the overall galaxy count distribution as a function of redshift for all three surveys (WALLABY, DINGO DEEP and UDEEP) in Fig.~\ref{fig:dndz_all} which
clearly demonstrates the `wedding cake' design of successfully deeper, overlapping, surveys which produce significant galaxy samples of order $10^2$
systems per redshift bin across the entire redshift range $z=0-0.43$ of ASKAP.

The WALLABY survey will be able to find $\sim 6 \times 10^{5}$ galaxies
at $S/N = 5$ which will be nearly two orders of magnitude more galaxies than the current \HI survey all-sky surveys have been able
to find. We have modelled the number counts as a function of $S/N$ in Table~\ref{tab:survey_results}, with a smooth drop
in galaxy detections for more demanding thresholds. DINGO may, through the DEEP and UDEEP tiered approach, detect $\sim 10^{5}$ galaxies over the
redshift range $0 - 0.43$, covering 5 billion years of cosmic evolution.

We also considered a Northern hemisphere \HI survey using the Westerbork array called WNSHS which was midway in flux sensitivity between WALLABY and 
DINGO DEEP. This facility can potentially detect $\sim 2 \,(5)\times 10^{5}$ galaxies dependent on the depth to which it surveys, making it 
an incredibly powerful dataset in its own right. However, if one were to combine
both WALLABY and WNSHS the potential science increases with a combined catalogue of over a million HI sources with integrated Signal-to-Noise of at least 5
with spectroscopic redshifts and a full $4\pi$ sky coverage. This will enable bulk flow studies and velocity field probes using a contiguous, blind dataset which
will dramatically improve the power of this promising probe of cosmology and structure formation.

\section{Conclusion} \label{sec:conclusion}
ASKAP is a uniquely powerful telescope, capable of cataloging $6 \times 10^{5}$ galaxies in the local Universe and probing the distribution of \HI
out to $z = 0.43$ with $5 \, - \, 6 \times 10^{4}$ galaxies (dependent on evolutionary possibilities), 
creating a more complete view of \HI than ever before. The systems we expect to find will probe
4 orders of magnitude in total halo mass and 7 orders of magnitude in stellar mass. 
This will greatly aid our understanding of galaxy formation as well as the nature of dark matter haloes. 
Furthermore a subsample of the nearby detections are well resolved by ASKAP, with $\sim 6000$ ($\sim 700$)  galaxies resolved by $>5$ ($>10$) 
beams, providing a valuable resource for the dynamical modelling of galaxies in 3D. However, if the full 6 km baseline for ASKAP is available
and the galaxies are surveyed with 10'' resolution then significantly more systems are well-resolved; $\sim 1.5 \times 10^{5}$ ($2 \times 10^{4}$) 
 galaxies resolved by $>5$ ($>10$) beams. 
 This resolved catalogue can then probe the small scale clustering of the dark matter and the influence that the baryons have at such scales~\citep[e.g.][]{Duffy:10}. 
 As well as a crucial aid to studying the velocity structure of nearby objects the postage-stamp zoom-in's 
 around galaxy detections are crucial to lowering the confusion rate of the distant galaxy population to sub-percent levels for all spectroscopic surveys
 with ASKAP. The postage-stamps will also be crucial in making possible unambiguous assignments of optical counterparts to the \HI detections, a key science case for
multi-wavelength surveys such as GAMA~\citep{Driver:11}.

As a community resource we make available catalogues for both surveys with the sky coordinates (RA and DEC), 
 redshift and observed redshift (i.e. including peculiar motion) as well as the stellar, halo, cold gas and \HI mass and finally the velocity width of the
 system. This catalogue will be invaluable in guiding survey preparations for Tully-Fisher studies, velocity field probes as well as correlation 
 function investigations to best utilise the formidable resource that ASKAP will represent for a wide range of astronomical fields.

\section*{Acknowledgements}
We would like to thank Martin Zwaan for his excellent comments in refereeing the paper. This work makes extensive use of the Coyote 
IDL libraries and AD is grateful to David Fanning in creating such a fantastic (free) resource. 
Additionally AD would like to thank Chris Power, Peder Norberg and Simon Driver for helpful discussions. 
DC acknowledges receipt of a QEII Fellowship. AD gratefully acknowledges the use of computer facilities purchased through a UWA 
Research Development Award. 
The Millennium Simulation used in this paper was carried out as part of the programme of the Virgo Consortium on the 
Regatta supercomputer of the Computing Centre of the Max-Planck-Society in Garching, and the Cosmology Machine supercomputer at 
the Institute for Computational Cosmology, Durham. The mock galaxy catalogues used in this work are available on request
and were generated using the Theoretical Astrophysical Observatory, see

\noindent \url{http://tao.it.swin.edu.au/mock-galaxy-factory}.
High resolution images and movies are available at

\noindent \url{http://ict.icrar.org/store/Movies/Duffy12c/}


\appendix
\section{HI Survey Lightcones}
\label{appendix}

\begin{figure*}
  \begin{center}
\epsfig{figure=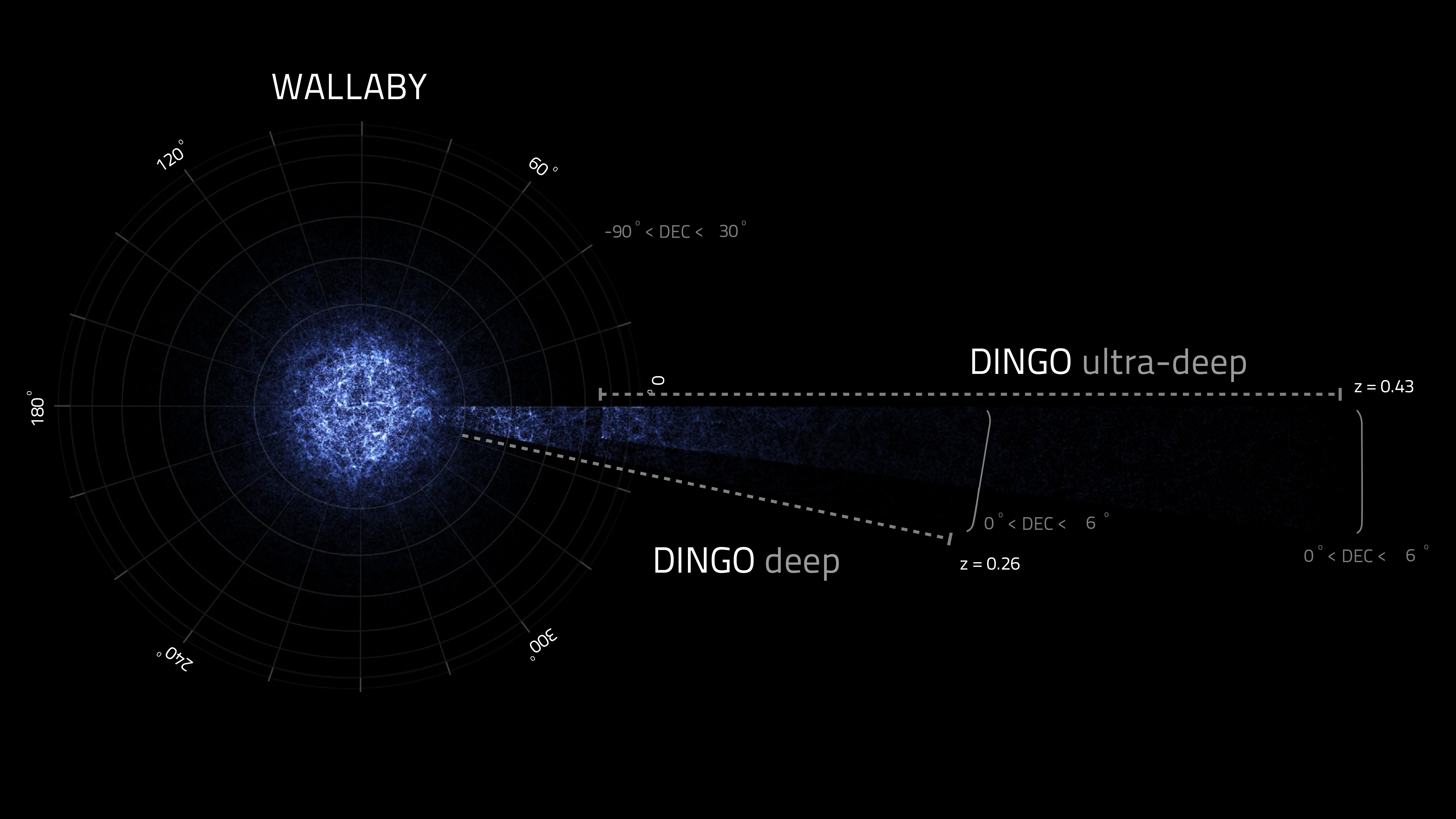, scale=0.3, angle=90}
    \caption[Survey]
            {The lightcone pie-plot for the shallow all-sky survey WALLABY and the two deeper surveys from DINGO. 
            We have taken the entire Declination range of the survey and 	
            projected it onto a redshift-RA 2D plot. Brightness is based on the number density
            of the sources in each map pixel. The cosmic web is clearly visible in this image, a key
            science driver of WALLABY is the measurement of the \HI mass function in different environments. In this volume are over
            $0.6$ million galaxy detections, which can be compared with HIPASS and ALFALFA which had approximately two orders 
            of magnitude fewer detections. The DINGO suvey will probe evolution in the high mass end of the \HI mass function
            over 4 billion years of cosmic time. Additionally, DINGO will overlap with existing {\sc GAMA} fields
            to enable a wealth of multi-wavelength data to be used when analysing the \HI detections; as well as enabling \HI spectral 
            stacking at given optical spectroscopic redshifts to extend the \HI detections. We reiterate that the DINGO fields are not actually
            contiguous as pictured here, but in fact are spaced across the sky. For the mock lightcone we made the simplifying 
            assumption that the fields were contiguous, this had no impact on the final number of galaxies predicted. 
            High res versions of this image
            and fly-through movies available at \url{http://ict.icrar.org/store/Movies/Duffy12c/}}
    \label{fig:lightcone}
  \end{center}
\end{figure*}

\label{lastpage}
\end{document}

%% file: jdefs.tex
\def\aj{AJ}					
\def\apj{ApJ}					
\def\apjl{ApJL}					
\def\apjs{ApJS}					
\def\aap{A\&A}					
\def\aaps{A\&AS}				
\def\mnras{MNRAS}				
\def\nat{Nature}				
\def\pasa{PASA}			
